\documentclass[a4paper,fleqn,usenatbib]{mnras}

\usepackage{mathptmx}

\usepackage[T1]{fontenc}
\usepackage{ae,aecompl}

\usepackage{graphicx}	
\usepackage{amsmath}	
\usepackage{amssymb}	

\title[RELHICs in $\Lambda$CDM]{The properties of ``dark'' $\Lambda$CDM halos in the Local Group}

\author[Ben\'itez-Llambay et al.]{Alejandro Ben\'itez-Llambay,$^{1}$\thanks{E-mail: alejandro.b.llambay@durham.ac.uk (ABL)}
Julio F. Navarro,$^{2}$, Carlos S. Frenk,${^1}$, 
\newauthor Till Sawala$^{3}$, Kyle Oman$^{2}$, Azadeh Fattahi$^{2}$, Matthieu Schaller$^{1}$,
\newauthor Joop Schaye$^{4}$, Robert A. Crain$^{5}$ and Tom Theuns$^{1}$.
\\
$^{1}$Institute for Computational Cosmology, Department of Physics, Durham University, South Road, Durham, DH1 3LE, UK \\
$^{2}$Department of Physics \& Astronomy, University of Victoria, BC, V8P 5C2, Canada \\
$^{3}$Department of Physics, University of Helsinki, Gustaf H\"allstr\"omin katu 2a, FI-00014, Helsinki, Finland \\
$^{4}$ Leiden Observatory, Leiden University, PO Box 9513, 2300 RA Leiden, The Netherlands\\
$^{5}$ Astrophysics Research Institute, Liverpool John Moores University, 146 Brownlow Hill, Liverpool, L3 5RF, UK\\}

\date{Accepted XXX. Received YYY; in original form ZZZ}

\pubyear{2016}

\begin{document}
\label{firstpage}
\pagerange{\pageref{firstpage}--\pageref{lastpage}}
\maketitle

\begin{abstract}
 We examine the baryon content of low-mass $\Lambda$CDM halos ($10^8<M_{200}/{\rm M_\odot}<5\times 10^{9}$) using the {\small APOSTLE} cosmological hydrodynamical simulations. Most of these systems are free of stars and have a gaseous content set by the combined effects of cosmic reionization, which imposes a mass-dependent upper limit, and of ram pressure stripping, which reduces it further in high-density regions. Halos mainly affected by reionization ({\small RELHICs}; REionization-Limited \ion{H}{I} Clouds) inhabit preferentially low-density regions and make up a population where the gas is in hydrostatic equilibrium with the dark matter potential and in thermal equilibrium with the ionizing UV background. Their thermodynamic properties are well specified, and their gas density and temperature profiles may be predicted in detail. Gas in {\small RELHICs} is nearly fully ionized but with neutral cores that span a large range of \ion{H}{I} masses and column densities and have negligible non-thermal broadening. We present predictions for their characteristic sizes and central column densities: the massive tail of the distribution should be within reach of future blind \ion{H}{I} surveys. Local Group {\small RELHICs} (LGRs) have some properties consistent with observed Ultra Compact High Velocity Clouds (UCHVCs) but the sheer number of the latter suggests that most UCHVCs are not {\small RELHICs}. Our results suggest that LGRs (i) should typically be beyond $500$ kpc from the Milky Way or M31; (ii) have positive Galactocentric radial velocities; (iii) \ion{H}{I} sizes not exceeding $1$ kpc, and (iv) should be nearly round. The detection and characterization of {\small RELHICs} would offer a unique probe of the small-scale clustering of cold dark matter.
\end{abstract}

\begin{keywords}
cosmology: theory -- (cosmology:) dark matter -- galaxies: halos -- (galaxies:) Local Group
\end{keywords}



\section{Introduction}
\label{SecIntro}

A defining prediction of hierarchically clustering models is that the Universe must be teeming with low-mass systems left over from the collapse of the early stages of the hierarchy \citep{White1978}. The $\Lambda$ cold dark matter ($\Lambda$CDM) paradigm is no exception; indeed, the abundance of $\Lambda$CDM halos massive enough, in principle, to host a galaxy is so high that they outnumber faint galaxies by a large factor \citep[see, e.g.,][]{Klypin1999,Moore1999}.  For example, more than $1,000$ halos with virial\footnote{We define virial quantities as those calculated within a radius where the mean inner density equals 200 times the critical density of the Universe, $\rho_{\rm crit}=3H^2/8\pi G$. Virial quantities are identified by a ``200'' subscript.} mass exceeding $10^8 \rm \ M_\odot$ are expected within $\sim 2 \rm \ Mpc$ from the barycentre of the Local Group (LG), a region that contains fewer than $100$ galaxies with baryonic masses exceeding $10^5 \rm \ M_\odot$ \citep[][and references therein]{Sawala2016}.

This discrepancy is usually explained by assuming that galaxies fail to form in halos below a certain halo mass, leaving a large number of systems essentially ``dark'', or free of stars.  The main culprit is cosmic reionization, which heats most baryons to $\sim 10^4 \rm \ K$ at relatively high redshift and prevents them from settling and condensing into galaxies in the shallow potential wells of low-mass halos\cite[e.g.,][]{Bullock2000}

The existence of these ``dark'' minihalos\footnote{Throughout this paper we shall refer to halos in the mass range $10^8<M_{200}/{\rm M_\odot}<5\times 10^{9}$ as minihalos.} is a cornerstone prediction of $\Lambda$CDM and their search has attracted great interest. Their presence could be inferred from their gravitational effects on dynamically cold structures, such as galaxy disks \citep[see][and references therein]{Feldmann2015} or thin stellar streams \citep{Ibata2002,Johnston2002,Carlberg2009}, or else from the distortions they may induce in gravitationally lensed images of distant galaxies \citep{Mao1998,Dalal2002,Vegetti2010,Hezaveh2016}. High-energy physicists, on the other hand, seek them as potential sources of energetic gamma rays powered by dark matter particle annihilation \citep{Diemand2007,Springel2008,Charles2016}.

A more prosaic alternative is to look for direct signatures of their baryonic content (which should be a nearly pristine H+He gaseous mix, given the lack of internal enrichment sources) in redshifted absorption against the light of luminous distant objects. Indeed, minihalos were once hypothesized as responsible for the forest of Lyman-$\alpha$ lines in the spectra of high redshift quasars \citep{Rees1986,Ikeuchi1986}, until it was realized that the large coherence length of absorption features was more naturally explained by the density ripples induced by CDM-driven fluctuations on larger scales \citep[see][for a review]{Rauch1998}.

``Dark'' minihalos might also be detectable in 21 cm emission and are actively sought in \ion{H}{I} surveys of the local Universe \citep[see, e.g.,][for a recent review]{Giovanelli2016}. Indeed, minihalos were proposed early on as hosts of the ``high velocity'' clouds (HVCs) of neutral hydrogen seen in 21 cm surveys of large areas of the sky \citep{Blitz1999,Braun1999}. 

The large sizes of HVCs, however, were shown to be incompatible with that interpretation: current models predict that gas in minihalos should be highly ionized by the cosmic UV background, except for a small central ``core'' of neutral hydrogen \citep{Sternberg2002}.  The mass and size of the neutral core depend sensitively on the mass of the halo and on the pressure of the surrounding medium. \ion{H}{I} cores of $\sim$ kpc size and mass $10^5$-$10^6 \rm \ M_\odot$ are expected in halos with virial mass in the $10^9$-$10^{10} \rm \ M_\odot$ range. At a putative distance of $\sim 1$ Mpc, these clouds would be much smaller and fainter than the typical HVC but still within range of current surveys \citep{Giovanelli2010}. 

The most promising minihalo candidates are the Ultra Compact High Velocity Clouds (UCHVCs) detected in surveys such as ALFALFA \citep{Adams2013} and GALFA \citep{Saul2012}. Their sizes and fluxes are consistent with minihalos in the Local Group volume, a result that has prompted deep follow-up imaging of some of the most prominent UCHVCs without obvious luminous counterparts in existing galaxy catalogs \citep[see, e.g.,][]{Sand2015,Bellazzini2015b}. These searches have revealed new dwarf galaxies, as illustrated by the discovery of Leo P, a gas-rich star forming dwarf at the edge of the Local Group \citep{Giovanelli2013}. 

\begin{figure}
	\includegraphics[width=\columnwidth]{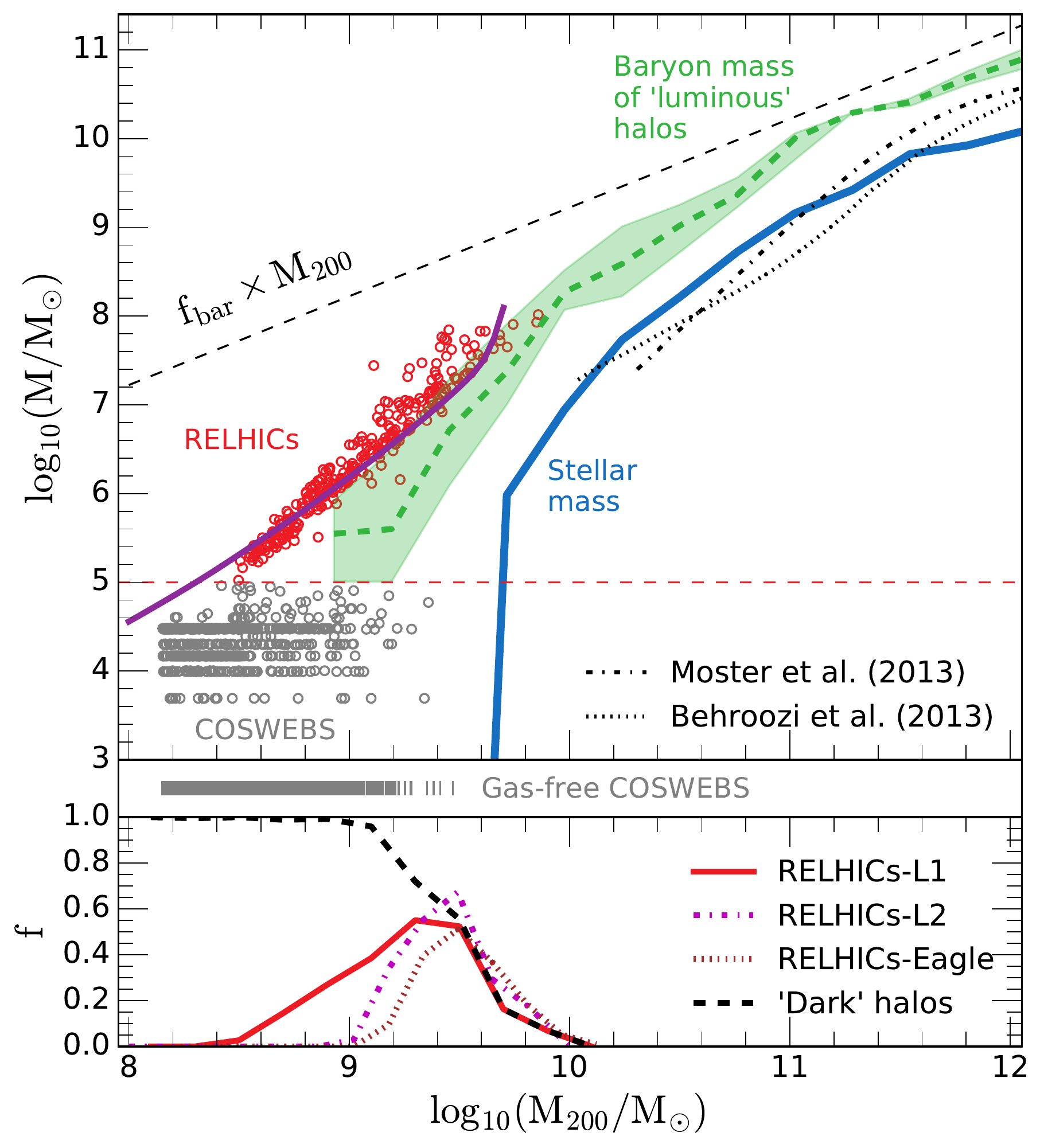}
    \caption{{\it Top:} Baryonic mass content of simulated halos at $z=0$ as a function of halo virial mass for the L1 simulations. The blue curve indicates the median stellar mass of simulated galaxies (measured within the galactic radius, $r_{\rm gal}$), whereas the green dashed curve shows the median total {\it bound} baryonic mass of the same galaxies, measured within the virial radius, $r_{200}$. The baryon mass of ``dark'' minihalos (i.e., star free) is shown with open circles; red for  {\small RELHICs} and grey for {\small COSWEBs}. The magenta solid line indicates the predicted gas mass of {\small RELHICs} according to the model of Appendix~\ref{SecApp}. For comparison, we also show the stellar mass vs halo mass relation derived using the abundance-matching technique by~\citet{Moster2013} (dot-dashed line) and~\citet{Behroozi2013} (dotted line). Gas-free minihalos are indicated in the middle panel. {\it Bottom:} Fraction of ``dark'' minihalos (thick dashed line). The other curves show the fraction of {\small RELHICs} computed for the resolution levels L1 (red solid) and L2 (purple dot-dashed), as well as for EAGLE run Recal-L025N0752 (brown dotted). Note the excellent agreement of all these simulations at the high-mass end.}
    \label{FigMasses}
\end{figure}

Some UCHVCs are thus clearly associated with faint galaxies, and, therefore, with minihalos. The converse, however, seems less clear. The sheer number of UCHVCs preclude many, if not most, of them from being associated with minihalos \citep{Garrison-Kimmel2014}, but it is unclear what criteria might be used to discriminate true minihalo candidates from \ion{H}{I} ``debris'' in the Galactic halo. 

\begin{figure*}
	\includegraphics[width=\textwidth]{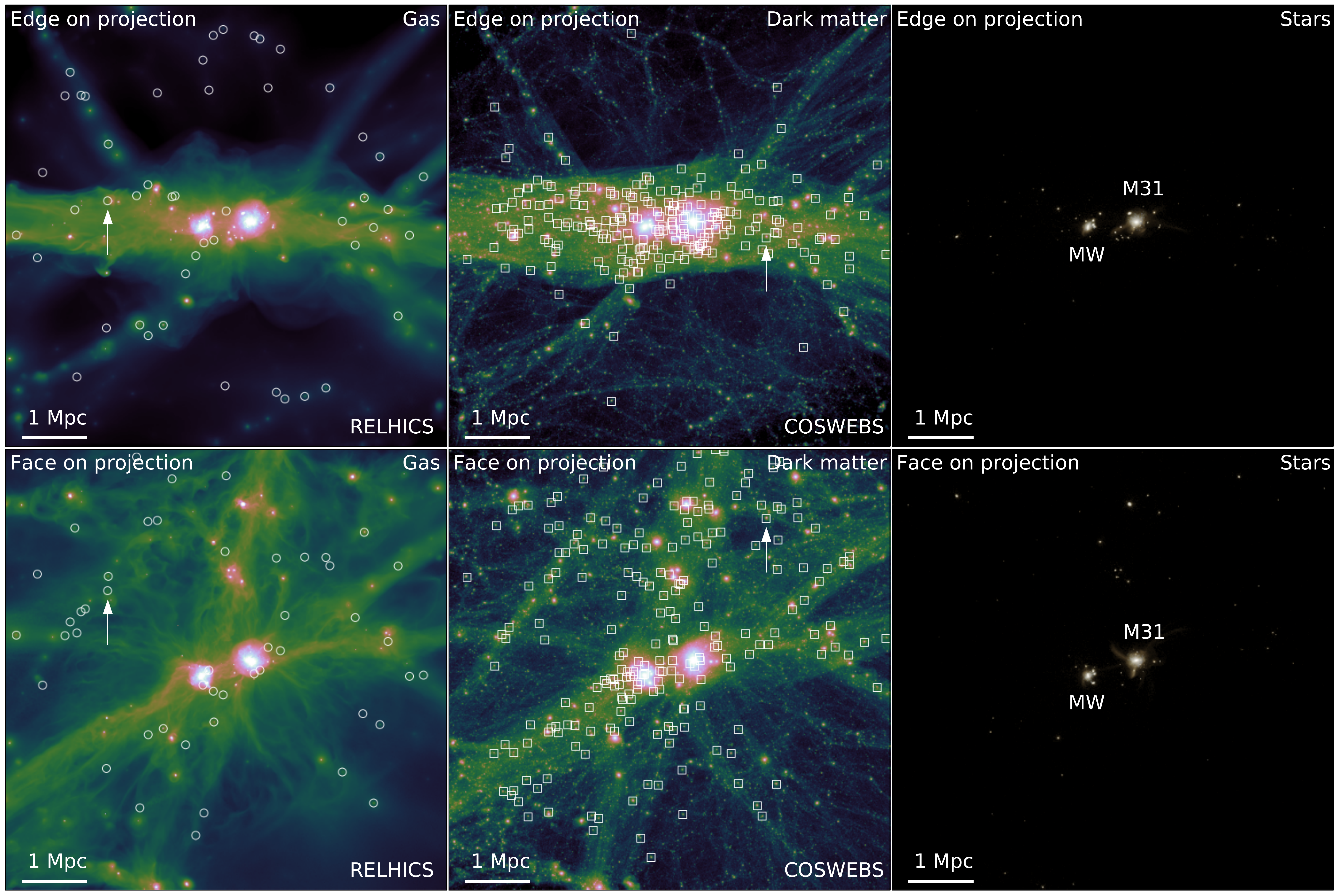}
    \caption{Panels, from left to right, show the distribution of gas, dark matter, and stars in one of the simulated volumes (namely V01-L1). Top and bottom rows show different orthogonal projections of a $7$ Mpc cubic box centred at the barycentre of the two main galaxies. Projections are chosen respect to the "sheet" that cross the volume. Colours indicate projected density, on a logarithmic scale. The location of {\small RELHICs} and {\small COSWEBs} are indicated in the left and middle panels, respectively. Note that {\small RELHICs} shun the high-density regions of the volume near the main galaxies, where cosmic web stripping is important and the population of {\small COSWEBs} dominates. Arrows show the positions of the individual {\small COSWEB} and {\small RELHIC} shown in Fig.~\ref{FigC} and Fig.~\ref{FigR}, respectively.}
    \label{FigSpatialDistribution}
\end{figure*}

We examine these issues here using the cosmological hydrodynamical simulations of the {\small APOSTLE}/EAGLE projects. We focus, in particular, on the gas content of ``dark'' minihalos. Given their lack of stars, and therefore of any energetic ``feedback'', two main mechanisms play a role in setting the gaseous content of minihalos: (i) cosmic reionization, which should evaporate much, but not all, of the baryons from such shallow potential wells, and (ii) ram pressure stripping by the cosmic web, which may unbind the gaseous content from minihalos that travel through dense filaments or ``pancakes'' of gas. \citet{Benitez-Llambay2013} show that the latter effect may reduce substantially the baryonic content of low-mass halos, especially in high-density regions such as groups of galaxies.

This paper is organized as follows. We begin in Section~\ref{SecSims} with a brief summary of the {\small APOSTLE}/EAGLE suite of simulations, followed in Section~\ref{SecRes} by a discussion of our main results. We analyse the baryon content of dark minihalos in Sec.~\ref{SecMgasM200}. We identify two populations of dark minihalos in Sec.~\ref{SecCR}; one where the properties of the gas component are mainly set by the ionizing background, and another where the gas has been nearly completely removed by cosmic web stripping. We discuss the properties of the former in Sec.~\ref{SecR}, where we also present a simple model that reproduces their main structural properties. We use this model to make predictions for their \ion{H}{I} content, column density profiles, and 21 cm line widths, and compare them with the properties of UCHVCs in Sec.~\ref{SecObs}. We conclude with a brief summary of our main conclusions in Sec.~\ref{SecConc}, and provide an appendix with an analytic model for the gas density and temperature profiles in minihalos.

\begin{figure*}
	\includegraphics[width=\textwidth]{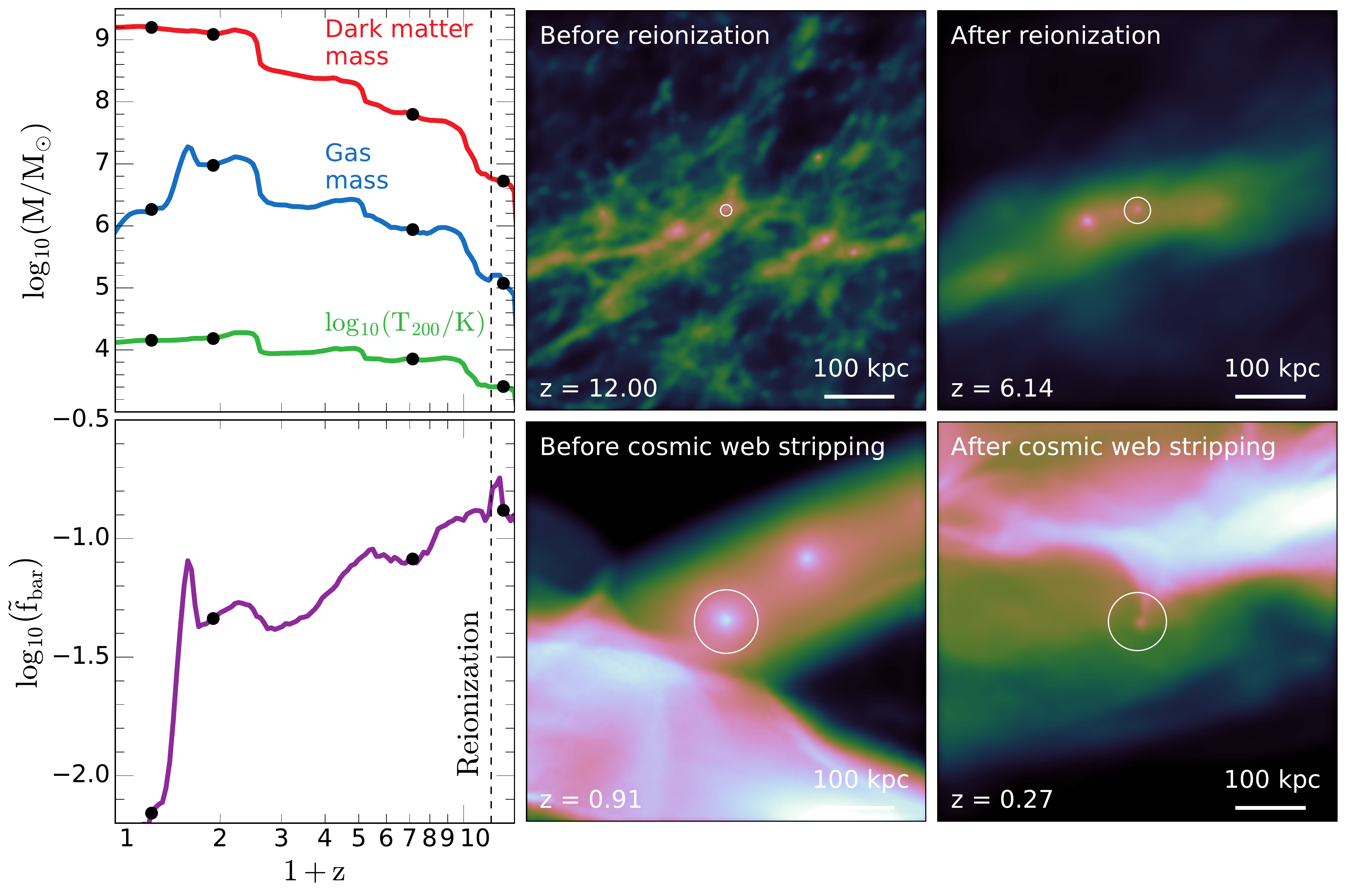}
    \caption{Evolution of a cosmic web stripped system from the V01-L1 simulations, with $ M_{200} \sim 10^{9.1} \rm M_{\odot}$ at redshift $z=0$. The left panels show the evolution of (from top to bottom) the dark mass and gas mass within $r_{200}$; the logarithm of the virial temperature, $T_{200} \sim 10^4 {\rm K} \left ( V_{200} / 17 {\rm km \ s^{-1} }\right )^2$, and (bottom panel) the gas mass fraction ($M_{\rm gas}/M_{200}$) in units of the universal value, $f_{\rm bar}=\Omega_b/(\Omega_0+\Omega_{\rm b})$. The four panels on the right show snapshots of the evolution, at the times indicated by the solid circles in the left panels. Note how reionization heats up the gas at early times, reducing the halo baryon content. In this particular example, the system loses essentially all of its bound mass after passing through a dense region of the cosmic web at $z\sim 0.6$. }
    \label{FigC} 
\end{figure*}

\section{Numerical Simulations}
\label{SecSims}

\subsection{The {\small APOSTLE} project}

We use a suite of cosmological hydrodynamical simulations from the {\small APOSTLE}\footnote{{\small APOSTLE} stands for ``A Project Of Simulating The Local Environment''} project~\citep{Fattahi2016a,Sawala2016}. These are zoom-in simulations that follow the formation of various Local Group-like cosmological environments. Twelve different realizations are followed at three different resolutions (L1, L2, and L3, in order of decreasing resolution). All volumes are selected from the DOVE N-body simulation~\citep{Jenkins2013} using the criteria described in~\cite{Fattahi2016a}. DOVE adopted a cosmological model with parameters consistent with WMAP7 measurements~\citep{Komatsu2011}, listed in Table~\ref{Tab:table1}.

\begin{table}
	\centering
	\caption{This table summarizes the main parameters of the simulations used in our analysis}
	\label{tab:example_table}
	\begin{tabular}{ccccr} 
		\hline
		$\rm H_{0}$ & $\rm \Omega_{0}$ & $\Omega_{b}$ & $\Omega_{\Lambda}$ & $z_{\rm reion}$\\
		\hline
		$70.4 \ \rm (km/s/Mpc) $ & $0.272$ & $0.0455$ & $0.728$ & $11.5$ \\
	\end{tabular}
	
	\begin{tabular}{ccccc} 
		\hline
		ID & $\rm \left (M_{gas}/\rm M_{\odot}\right )$ & $\rm \left ( M_{drk}/\rm M_{\odot} \right )$ & $\rm n_{part}$ & $(\epsilon_0/\rm pc)$ \\
		\hline
		$\rm V01-L1$ & $9.89 \times 10^3$ & $4.92 \times 10^4$ & $\sim 2.6 \times 10^8$ & $134$\\
		$\rm V04-L1$ & $4.93 \times 10^3$ & $2.45 \times 10^4$ & $\sim 5.4 \times 10^8$ & $134$\\
		$\rm V11-L1$ & $1.01 \times 10^4$ & $5.02 \times 10^4$ & $\sim 2.4 \times 10^8$ & $134$\\
		
		\hline		
	\end{tabular}
	\label{Tab:table1}
\end{table}

Each simulated volume contains a relatively isolated pair of halos of combined virial mass in the range $10^{12.2}$-$10^{12.5} \,\rm M_{\odot}$; the two halos (meant to represent the Milky Way and M31) have separations $\sim 600$-$1000 \rm \ kpc$, approach with a relative radial velocity $-250$-$0$ km s$^{-1}$, and have tangential velocities that do not exceed $100$ km s$^{-1}$~\cite[see ][]{Fattahi2016a}.

Different resolution levels are chosen so that each level improves on the previous by a factor of $\sim 12$ in particle mass, and a factor of $\sim 12^{1/3}$ in gravitational force resolution. The highest-resolution level (L1) has a baryon particle mass of $(5 - 10) \times 10^3 \rm \ M_\odot$ (dark matter particles are $\sim 5$ times heavier), and a Plummer-equivalent gravitational softening of $\epsilon_0 = 134$ pc. At the time of writing, all twelve volumes have been completed at L3 and L2 resolution but only three volumes (V01, V04, and V11) have also been completed at L1 resolution. We focus on those three volumes in the rest of this paper, although we use results from lower resolution runs to assess the sensitivity of our results to numerical resolution.

\begin{figure*}
	\includegraphics[width=\textwidth]{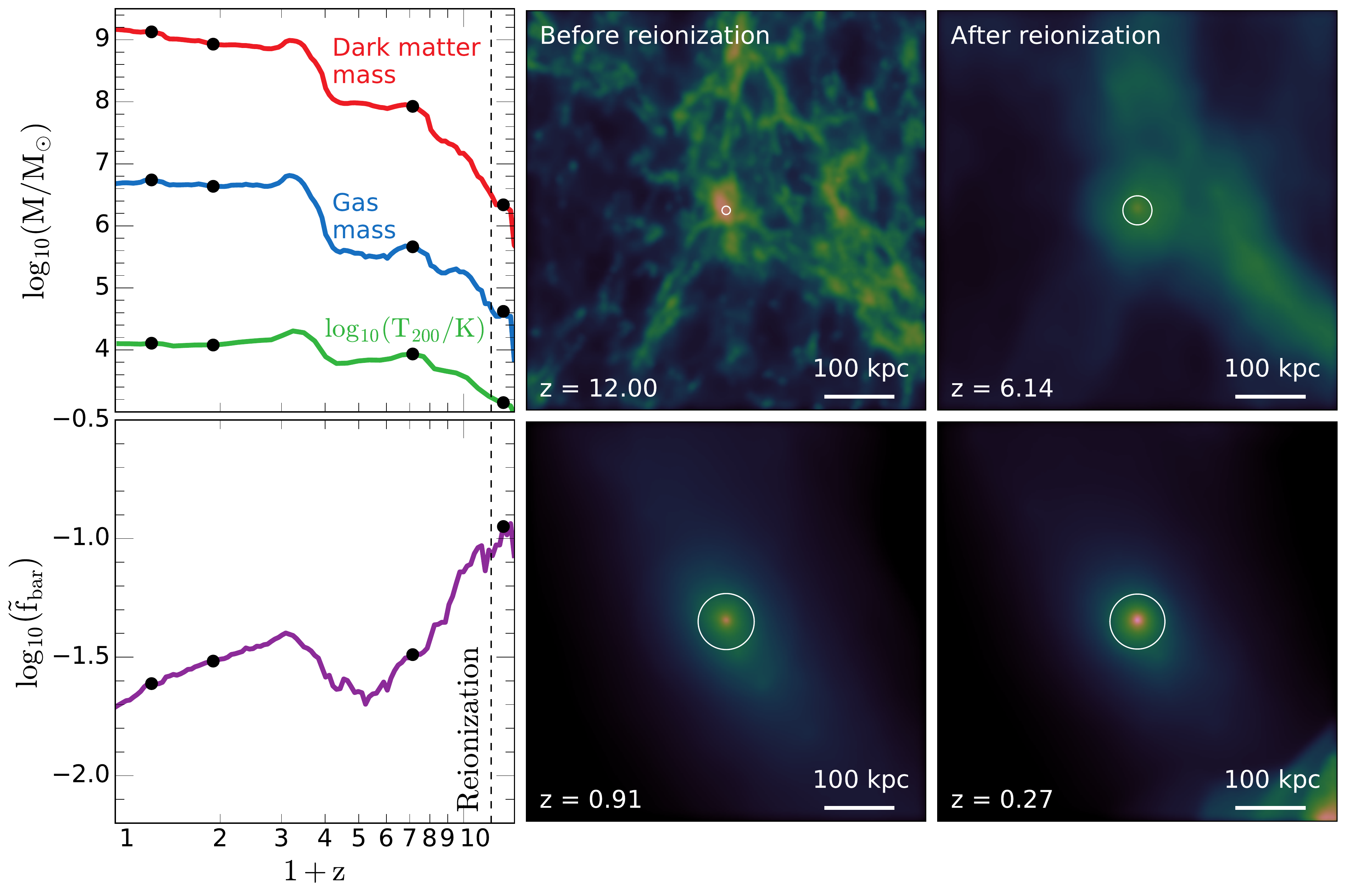}
    \caption{As Fig.~\ref{FigC}, but for the case of {\small RELHIC} with $M_{200} \sim 10^{9.1} \rm \ M_{\odot}$ . This halo has its baryon content substantially reduced by cosmic reionization, but it is not affected by cosmic web stripping, since it inhabits the low density outskirts of the Local Group volume. Its baryon content is essentially constant after $z\sim 5$-$6$ and is set by the hydrostatic balance of UV-heated gas in the potential of the minihalo. (see Appendix~\ref{SecApp}).}
    \label{FigR}
\end{figure*}

{\small APOSTLE} simulations were run with the same modified version of the {\tt P-Gadget3} code, last described by ~\cite{Springel2005}, which was used to run the simulations of the {\tt EAGLE} project~\citep{Schaye2015, Crain2015}
The {\tt EAGLE} code includes a set of subgrid prescriptions to account for the effects of radiative cooling, photoheating, star formation, energy feedback from star formation, and AGN feedback, among others. The adjustable numerical parameters of the subgrid modules were chosen to provide an approximate match to the galaxy stellar mass function and galaxy sizes over cosmological volumes, and correspond to that of the EAGLE 'Reference' runs.  As discussed by \citet{Sawala2016} the three {\small APOSTLE} resolutions yield very similar galaxy stellar mass functions (over the range resolved with more than 50 star particles), indicating good numerical convergence in the dwarf galaxy regime probed by {\small APOSTLE}. In addition, the {\small APOSTLE} galaxy mass-halo mass relation matches rather well the abundance-matching constraints for galaxies in the range $10^7 \le M_{\rm str}/M_{\odot} \le 10^{10}$ (see Fig.~\ref{FigMasses}). We conclude that, at least in the low-mass regime, our results are weakly sensitive to numerical resolution, and we therefore attempt no further parameter recalibration.


\subsection{Reionization and the UV background}

Cooling and photoheating processes in the {\tt EAGLE} code are implemented following the procedures outlined in~\cite{Wiersma2009a}. In brief, the thermodynamic state of the gas is modelled using {\tt CLOUDY}~\citep{Ferland1998}, assuming ionization equilibrium with the cosmic microwave background (CMB) and a  spatially uniform evolving UV/X-ray background radiation field as calculated by ~\cite{Haardt2001} (HM01 thereafter) in the optically thin regime. The simulation neglects self-shielding which could have an impact on regions with $n_{\rm H} > 10^{-2} \rm \ cm^{-2}$, and temperatures below $\sim 10^4 \rm \ K$~\citep[see, e.g,][]{Schaye2001,Rahmati2013}\footnote{Our study focuses on systems with gas densities $n_{\rm H} \le 10^{-1} \rm \ cm^{-2}$ and temperatures $T \ge 10^4 \rm \ K$. Self-shielding could in principle change the properties of some of these systems since the temperature would be slightly reduced, thus increasing the \ion{H}{I} mass predicted in Sec.~\ref{SecHI}.}.
Reionization of the Universe is modelled by switching on the HM01 background radiation field at redshift $z_{\rm reion} = 11.5$. The photoheating and photoionizing rates are kept fixed in the redshift range $z=9-11.5$. For redshift $z \le 9$, the UV background is allowed to evolve, and reaches a maximum at redshift $z \sim 2$. In addition, an extra heating of 2eV per proton mass is injected to the gas particles at $z_{\rm reion}$, which accounts for a boost in the photoheating rates during reionization relative to the optically thin rates assumed here, ensuring that the photoionized gas is rapidly heated to a temperature of $\sim 10^4 \rm \ K$. This is done instantaneously for \ion{H}, but for \ion{He}{II} the extra heat is distributed in redshift with a Gaussian of width $0.5$, centred at $z=3.5$. 

For redshift $z>11.5$, the net cooling\footnote{We refer to the net cooling of the gas as the difference between the radiative heating and cooling processes.} of the gas is computed by exposing it to the CMB and the photodissociating background obtained by cutting the HM01 spectrum at 1 Ryd. Note that the presence of photodissociating radiation and the finite resolution of our simulations imply that we cannot model the formation of Pop III stars via $\rm H_{2}$ cooling in minihalos, which could play a role for isolated halos of mass $\sim 10^5 \rm \ M_{\odot}$.

\subsection{Halo finding}

Halos are identified in the simulations using the group finder {\tt SUBFIND}~\citep{Springel2001, Dolag2009}, which identifies self-bound substructures within a catalogue of friends-of-friends (FoF) halos built with a linking length of 0.2 times the mean interparticle separation.  {\tt SUBFIND} provides a list of  self-bound subhalos within each FoF halo, organized as a ``central'' halo and its respective ``satellites''. 


Most of our analysis is based on central halos identified at redshift $z=0$ within a spherical volume of radius $3.5 \rm \ Mpc$ centred at the barycentre of each simulated ``Local Group''. 
We keep for analysis all central halos with $M_{200} \ge 10^8 \ \rm M_{\odot}$ (i.e., with typically more than $3000$ dark matter particles). These limits in volume and mass ensure that all selected halos are far enough from the boundaries of the high-resolution zoom-in region and that we are able to resolve them confidently. 

\section{Results}
\label{SecRes}

\subsection{Baryonic content of {\small APOSTLE} halos}
\label{SecMgasM200}

\begin{figure}	
\includegraphics[width=\columnwidth]{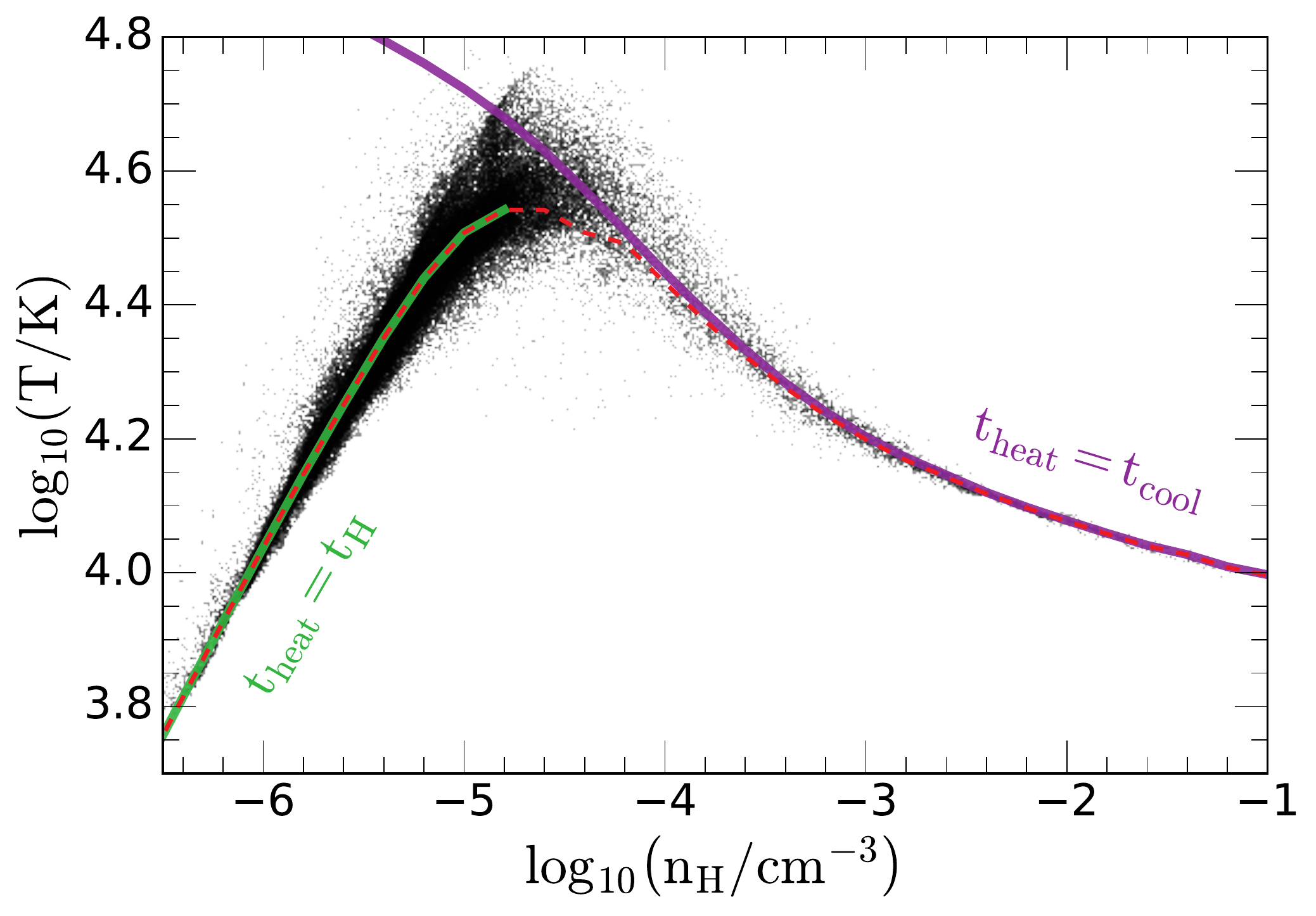}
\caption{Temperature-density diagram for all gas particles bound to {\small RELHICs}. All particles of all {\small RELHICs} are shown, and compared with two curves indicating (i) the loci of particles where the photoheating timescale equals the age of the Universe, $t_{H} \approx 13.76 \rm \ Gyr$ (green dashed curve); and (ii) the loci where photoheating and radiative cooling are in equilibrium (thick magenta curve). Gas in {\small RELHICs} has been photoheated to a density-dependent temperature that matches one of those two regimes. The tight relation between density and temperature that results defines a temperature-density relation for gas in {\small RELHICs} (red dashed line) that can be used to derive density and temperature profiles assuming that the gas is in hydrostatic equilibrium to the potential well of the dark matter.}
    \label{FigRhoT}
\end{figure}

We begin by analysing the baryonic content within the virial boundaries of the simulated halos. This is presented in the top panel of Figure~\ref{FigMasses}, where we show the relation between virial mass and the mass of various baryonic components for the three ``high-resolution'' volumes. The oblique dashed line indicates, for reference, the theoretical maximum baryonic mass within the virial radius, $M_{\rm bar}=f_{\rm bar}M_{200}$, where $f_{\rm bar}=\Omega_{\rm b}/(\Omega_{\rm 0}+\Omega_{\rm b}=0.167$ is the universal baryon fraction.

The blue solid line indicates the median stellar mass "bound" to the central galaxies. Note that we only consider ``central'' galaxies in this figure; i.e., the most massive subhalo of each FoF halo. 

The median stellar mass plummets below a virial mass of $\sim 10^{10} \rm \ M_\odot$, mainly because not all those low-mass halos harbour luminous galaxies. This may be seen from the thick black dashed line in the bottom panel of Fig.~\ref{FigMasses}, which shows the fraction of central galaxies that do not have stars. All halos above $10^{10} \rm \ M_\odot$ have luminous galaxies, but the fraction dips to $50\%$ at $M_{200} \sim 5 \times 10^9 \rm \ M_\odot$. Below $10^9 \rm \ M_\odot$ essentially all halos are ``dark''\footnote{Strictly speaking, these halos have galaxies less massive than a few $10^3 \rm \ M_\odot$ in stars, the mass of one baryon particle at this resolution level.}. 

The median total baryon mass bound to {\it luminous} halos, measured within $r_{200}$, is shown by the green curve in Fig.~\ref{FigMasses}, with a shading that indicates $\pm1\sigma$ dispersion. The total gas mass within $r_{200}$ bound\footnote{We note that {\tt SUBFIND} can at times underestimate these masses, especially in regions where the mean ambient gas density is comparable to the density in the outer regions of the minihalo or when a minihalo is embedded in hotter gas; the masses shown here have been carefully recomputed to take those issues into account.} to ``dark'' (i.e., star free) halos is shown with open circles. Two populations are clearly apparent: one where the bound gas mass is so small ($<10^5 \rm \  M_\odot$, shown in grey) that it can barely be measured (the gas particle mass is $\sim 10^4 \rm \ M_\odot$ at resolution level L1); and another where the gas mass correlates tightly with virial mass (shown in red). We shall hereafter refer to the former as ``{\small COSWEBs}'' (short for cosmic web-stripped halos) and to the latter as ``{\small RELHICs}'' (for Reionization-Limited \ion{H}{I} Clouds). 


Before discussing the origin of these two populations, we show in the bottom panel of Fig.~\ref{FigMasses} their relative fractions as a function of mass, as well as the dependence on numerical resolution. The thick solid red curve indicates the fraction of {\small RELHICs} in the highest-resolution (L1 level) runs: {\small RELHICs} inhabit halos spanning a small range in virial mass, $3\times 10^{8}<M_{200}/\rm M_\odot<5\times 10^{9}$, making up about half of all systems at the mid point of that range. Lower-resolution simulations (L2 runs, shown with a dot-dashed line), as well as simulations of a larger volume\footnote{EAGLE run Recal-L025N0752.} at L2 resolution yield {\small RELHIC} fractions similar to that of L1 runs at the high-mass end of the range, but underestimate their abundance at the low-mass end, below a virial mass of $\sim 2\times 10^9 \rm \ M_\odot$.  The drop in {\small RELHIC} fractions below that mass is mainly a result of limited numerical resolution. As we discuss below, the main difference between {\small RELHICs} and {\small COSWEBs} is environmental, so we would expect the ratio of the two to approach a constant at low virial masses.  Interestingly, although not shown here, {\small RELHICs} in all of these runs track the same gas mass-virial mass trend shown in the upper panel of Fig.~\ref{FigMasses}, regardless of resolution.

\begin{figure*}
	\includegraphics[width=\textwidth]{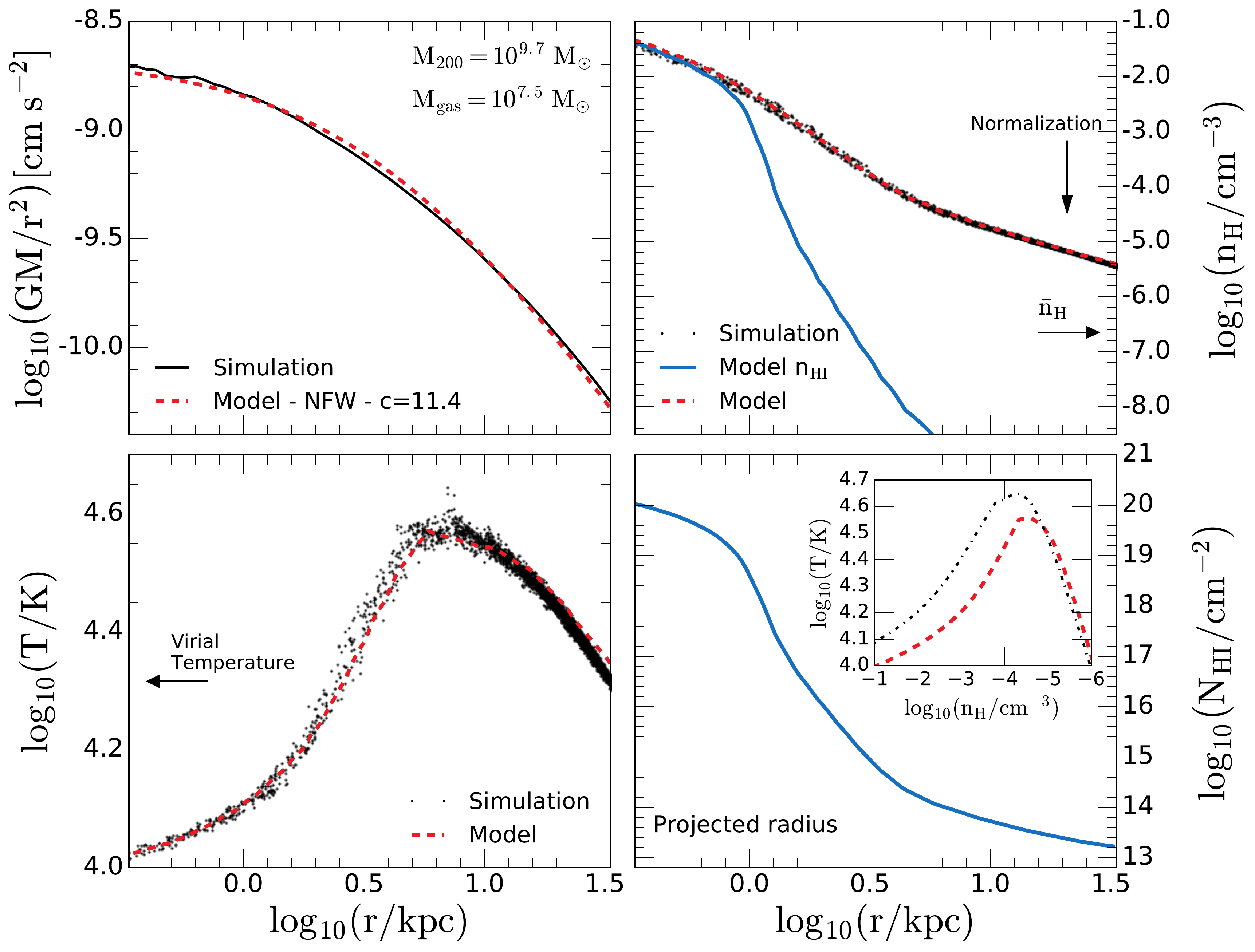}
        \caption{Acceleration (top left), temperature (bottom left) and gas density (top right) profiles for one {\small RELHIC} of virial mass $\rm M_{200} \sim 5\times 10^{9} \rm \ M_{\odot}$ at redshift $z=0$. Points indicate individual particles bound to the {\small RELHIC}. The dashed red line in the top left panel is an NFW fit to the spherically averaged acceleration profile of the halo ($a(r)=GM(r)/r^2$). Dashed curves in the other two panels are results of the model presented in Appendix~\ref{SecApp}, where the gas is simply assumed to be in hydrostatic equilibrium. The inset in the bottom right panel shows, with red-dashed line, the temperature-density relation (Fig.~\ref{FigRhoT}) used in the model (note that we have inverted the x-axis respect to Fig.~\ref{FigRhoT} for clarity). For comparison, we also show with a black dot-dashed line the temperature-density relation expected from a stronger UV background, which in this case corresponds to the $z=1$ HM01 spectrum. Blue curves in the right hand panels show the neutral hydrogen profiles, number density in the top, and column density in the bottom.}
	\label{FigRadProf}
\end{figure*}

\subsection{COSWEBs and RELHICs}
\label{SecCR}

The spatial distribution of the two populations ({\small RELHICs} and {\small COSWEBs}) are clearly different, as shown in Fig.~\ref{FigSpatialDistribution}, hinting at an environmental origin of their distinction. This figure shows two orthogonal projections of one {\small APOSTLE} volume run at the highest-resolution (V01-L1): the two main LG galaxies are clearly visible near the centre, surrounded by a two-dimensional ``sheet'' of gas that extends out to several Mpc from the LG barycentre. {\small COSWEBs}, shown as open squares in the middle panels, cluster around the main galaxies and inhabit the mid plane of the sheet, whereas {\small RELHICs} (open circles in the left panels) populate underdense regions in the periphery of the LG volume.

The evolution of a typical {\small COSWEB} is shown in Fig.~\ref{FigC}. The top left panel shows the evolution of the various mass components, measured within the virial radius of the most massive progenitor. As expected, the dark mass grows monotonically, with occasional jumps corresponding to merger events. The gas mass largely follows suit, except at late times, when it drops dramatically, in this example at $z\sim 0.6$. The drop is more clearly seen in the bottom left panel, which tracks $\tilde f_{\rm bar}$, the baryon fraction of the halo expressed in units of the universal baryon fraction.

\begin{figure}
	\includegraphics[scale=0.5]{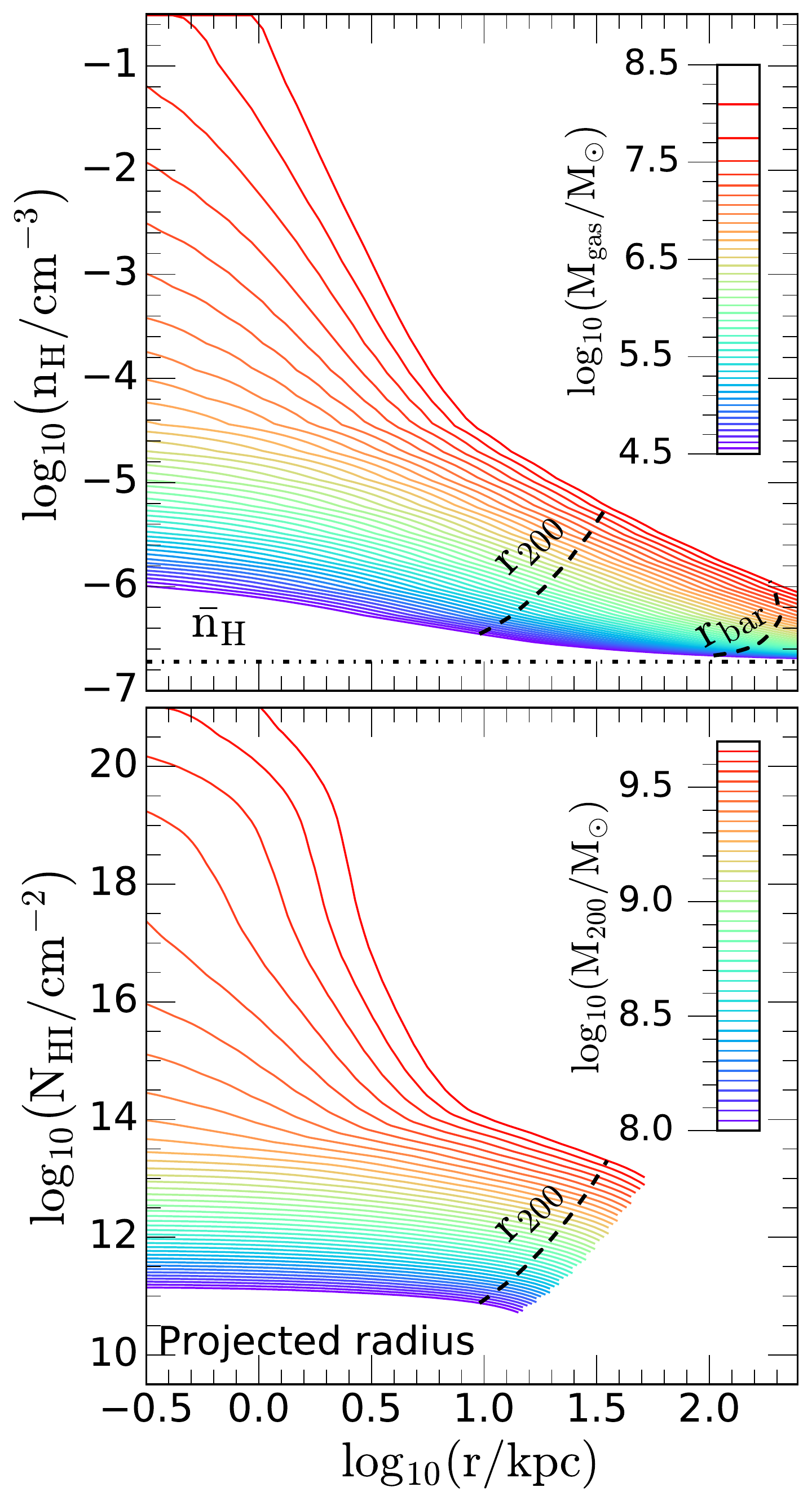}
    \caption{Model gas density profiles for {\small RELHICs}. The model, which is described in detail in Appendix~\ref{SecApp}, solves the equations of hydrostatic equilibrium in NFW halos with the assumption that gas follows the temperature-density relation shown in Fig.~\ref{FigRhoT} and the boundary condition that densities approach the mean density of the Universe (${\bar n}_{\rm H}$) at $r\gg r_{200}$. The top panel shows gas densities, and the bottom panel column densities of \ion{H}{I}. Neutral fractions are computed as in the appendix A1 of \citet{Rahmati2013}. Two characteristic radii are indicated in the top panel; the virial radius of each halo, $r_{200}$, and the radius where the total enclosed gas mass equals $f_{\rm bar}\, M_{200}$, with $f_{\rm bar}$ the universal baryon fraction. Curves in each panel correspond to the same halos, from top to bottom, and are coloured by either gas mass (top panel) or virial mass (bottom panel).}
    \label{FigModProf}
\end{figure}

The four panels on the right of Fig.~\ref{FigC} show snapshots of the gas component at different stages of the evolution, at times marked by the solid circles in the left panels. The top two right panels show the {\small COSWEB} just before and after reionization, when the gas is suddenly heated to $\sim 10^4$K, a temperature that exceeds the virial temperature of the halo at that time (see the bottom green curve, which tracks $\log_{10} T_{\rm 200}/$K). The gas is therefore too hot to remain bound, and evaporates from the potential well of the halo. By $z\sim 2$, the halo has only been able to retain about $1/20$th of its baryons.  Just after $z \sim 1$, however, the {\small COSWEB} travels through an overdense region of the cosmic web, which ram-pressure strips the remainder of the gas, reducing its bound gas content to a negligible amount. The {\small COSWEB} is unable to re-accrete gas by redshift $z=0$, mainly because it is now moving relatively fast within the much hotter LG environment. {\small COSWEBs} are thus star/gas-free minihalos whose gaseous content has been removed by the cosmic web~\citep[see][for a more extensive discussion of cosmic web stripping]{Benitez-Llambay2013}.

By contrast, {\small RELHICs} are minihalos that have avoided dense regions of the cosmic web, and have therefore been able to retain more baryons. The evolution of a typical {\small RELHIC} is shown in Fig.~\ref{FigR}. Comparing this to that of the {\small COSWEB} in Fig.~\ref{FigC}, we see that the main difference is that the {\small RELHIC} baryon fraction remains more or less constant after the initial drop caused by reionization. At $z=0$ this particular {\small RELHIC} has been able to retain about $1/30$th of the universal baryon fraction.  What sets the final baryon content of {\small RELHICs} and, in particular, what is the origin of the strong correlation between bound gas mass and virial mass shown in Fig.~\ref{FigMasses}? We address these questions next.

\section{RELHICs}
\label{SecR}

\subsection{Gas densities and temperatures}
\label{SecRhoT}

We start by considering the thermodynamic state of the gas in {\small RELHICs}. This is shown in Fig.~\ref{FigRhoT}, where we plot the density and temperature of all gas particles bound to the 249 {\small RELHICs} identified in all three simulated volumes. We have excluded 9 {\small RELHICs} (i.e., $3$ per volume) with masses $M_{200} \gtrsim 10^{9.7} \rm M_{\odot}$, whose central gas densities reach values $n_{\rm H} \approx 10^{-1} \rm \ cm^{-3}$, and thus their central thermodynamic properties are expected to be governed by the effective equation of state imposed in the EAGLE code on the unresolved multiphase interstellar/star forming medium. 

The gas in the remaining {\small RELHICs} spans nearly seven decades in density, from the mean baryon density of the Universe (${\bar n_{\rm H}}\approx 10^{-6.7} \rm \ cm^{-3}$) to densities 100 times below the threshold chosen for star formation in EAGLE ($n_{\rm H, th}\approx 10 \,$cm$^{-3}$ for a gas with primordial composition)\footnote{We discuss the impact of the particular choice of the star formation density threshold on our results in Appendix~\ref{SecApp2}.}. Gas temperatures are a non-monotonic function of density, first climbing to a maximum of $\sim 4\times 10^4 \rm \ K$ at $n_{\rm H} \sim 10^{-4.8} \rm \ cm^{-3}$, and then dropping gradually at higher densities, approaching $10^4 \rm \ K$.

The gas is, on average, hotter than the virial temperature of a typical {\small RELHIC} ($T_{200}=10^{4.1} \rm \ K$ for a virial mass of $2\times 10^9 \rm \ M_\odot$), implying that the gas temperature is largely set by the ionizing background, and not by the gravitational collapse of the halo. This is further demonstrated by the green dashed curve, which indicates where the photoionizing heating time-scale equals the age of the Universe, $t_{H} \approx 13.76 \rm \ Gyr$. Temperatures of low density gas in {\small RELHICs} are clearly set by the ionizing background. 

At densities $> 10^{-4.8}$cm$^{-3}$ radiative cooling induced by collisional effects becomes more important and the gas settles at the ``equilibrium'' temperatures where radiative cooling effects balance photoheating from the ionizing background, shown by the thick purple line in Fig.~\ref{FigRhoT}~\citep[see e.g.,][]{Haehnelt1996, Theuns1998}. As is clear from this figure, these two regimes describe very well the temperature-density relation of gas in {\small RELHICs} (shown by the red dashed curve).

\subsection{Gas masses}
\label{SecGasM}

The $n_{\rm H}$-$T$ relation followed by gas particles in {\small RELHICs} effectively defines a pressure-density relation, $P=P(\rho)$, that enables us to estimate the gas mass bound to a halo of given virial mass. This may be done by assuming that the gas is in hydrostatic equilibrium within the potential of the dark halo and solving: 
\begin{equation}
\displaystyle\frac{1}{\rho}\frac{dP}{dr} = -\frac{GM(r)}{r^2},
\label{EqHydroEquil}
\end{equation}
to give a density profile that may be integrated to compute the total gas mass within $r_{200}$, once a boundary condition (e.g., an external pressure) is chosen. 

The simplest choice is to assume that far from the virial boundary the gas reaches the mean baryon density of the Universe, at the appropriate temperature set by the ionizing background: this specifies the external pressure that closes the set of equations, enabling a simple estimate of {\small RELHIC} gas masses. 

We present details of the calculation in Appendix~\ref{SecApp} and show the main result by the thick purple line in Fig.~\ref{FigMasses}. Despite its simplicity, the model predicts accurately the gas mass of {\small RELHICs} for halos not exceeding virial masses of order $5\times 10^9 \rm \ M_\odot$. At higher masses gravitational heating becomes important and, in addition, the central densities become high enough for self-shielding and cooling processes to become important; in those halos the gas would not be able to stay in hydrostatic equilibrium, but will collapse into a rotationally supported disk where it may form stars. Indeed, very few, if any, halos above $5\times 10^9 \rm \ M_\odot$ remain ``dark'', as shown in the bottom panel of Fig.~\ref{FigMasses}.

\begin{figure*}
	\includegraphics[width=\textwidth]{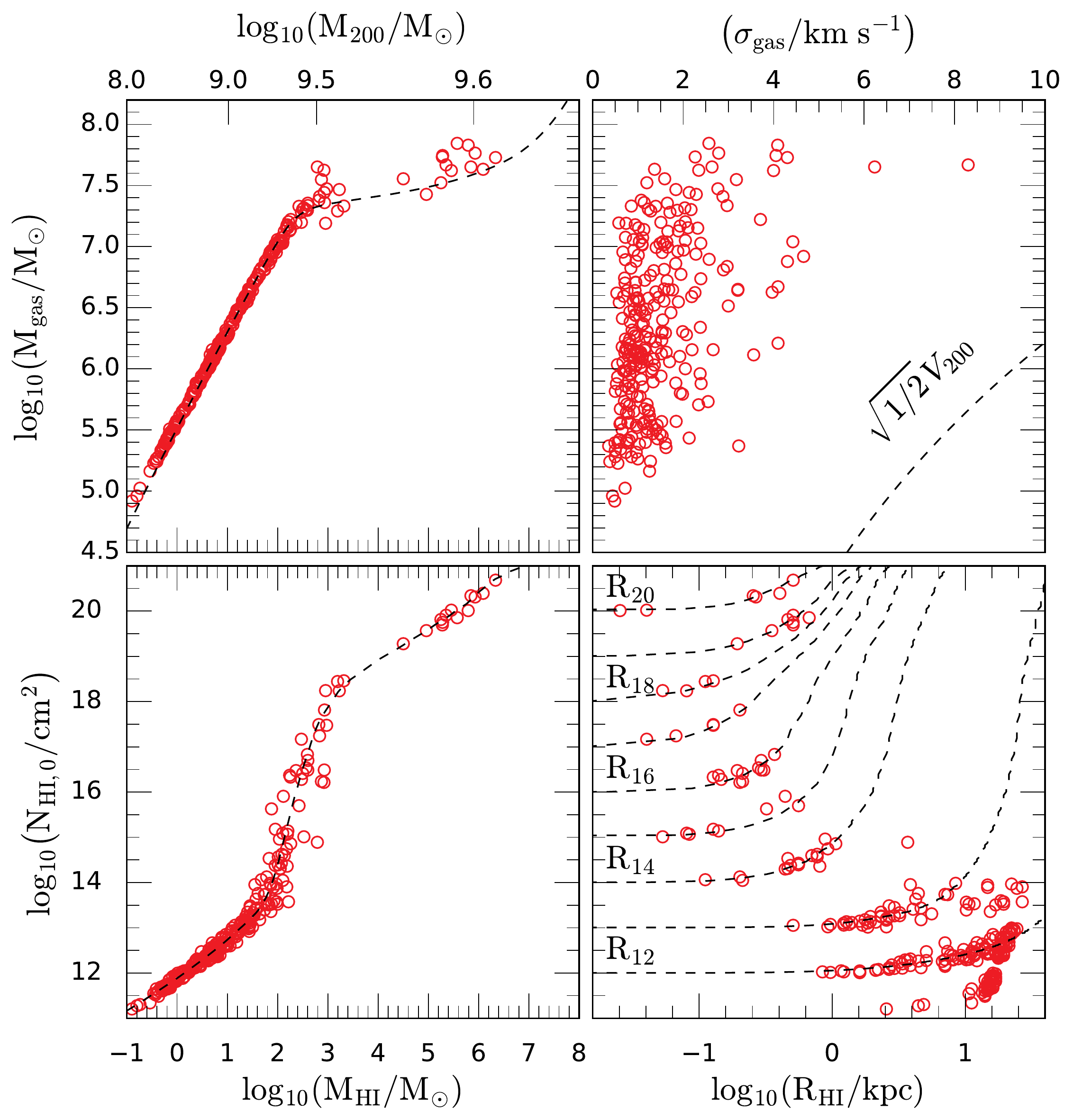}
    \caption{Properties of {\small RELHICs} (open circles), compared with those of the model presented in Appendix~\ref{SecApp}. Left panels show, as a function of \ion{H}{I} mass within $r_{200}$, the total gas mass (top) and the central \ion{H}{I} column density $\rm N_{\ion{H}{I},0}$. Right panels show the total gas mass vs the velocity dispersion in bulk motions of the gas (top); and the central \ion{H}{I} column density $\rm N_{\ion{H}{I},0}$ vs \ion{H}{I} size, $R_{\rm HI}$. Several characteristic radii for the latter are shown; from top to bottom the dashed lines indicate the radius of the iso column density contour of $10^{20}$, $10^{19}$, etc, in units of cm$^{-2}$. Each {\small RELHIC} is shown at the column density immediately  below its central value; for example, the radius of the $10^{18}$ cm$^{-2}$ contour is shown for  those {\small RELHICs} with central column densities in the range $10^{18}<N_{\ion{H}{I},0}/$cm$^{-2}<10^{19}$, and so on.}
    \label{FigHI}
\end{figure*}

\subsection{Gas and temperature profiles}
\label{SecRhoTvsR}

We may test further the simple hydrostatic model described in Appendix~\ref{SecApp} by using it to predict the density and temperature profiles of the gas component of {\small RELHICs} and comparing them with the simulation results. Figure~\ref{FigRadProf} shows an example for a relatively massive {\small RELHIC}; $M_{200} \sim 5\times 10^{9} \rm \ M_{\odot}$ and $\rm M_{gas} \sim 3\times 10^{7} \rm \ M_{\odot}$. 

We first measure the acceleration profile of the halo, assuming spherical symmetry: $a(r)=GM(r)/r^2$, where $M(r)$ is the total enclosed mass within radius $r$. Baryons contribute so little mass that this is effectively equivalent to the dark matter acceleration profile. We show $a(r)$ in the top left panel of Fig.~\ref{FigRadProf} by the solid black line. The red dashed curve is a fit to this acceleration profile, namely, a Navarro-Frenk-White \citep[][hereafter NFW]{Navarro1996,Navarro1997} profile with concentration parameter $c=11.4$. 

We then integrate Eq.~\ref{EqHydroEquil} numerically, using a fit to the temperature-density relation (see the red dashed curve in the inset of the bottom right panel of Fig.~\ref{FigRadProf}), and normalizing the profile at the radius that contains half of all gas particles\footnote{This normalization procedure improves fits to individual halos, but its results are not very different from those obtained assuming the simple boundary condition that the gas profile should converge to the mean density of the Universe at large radii.}. The result for the density profile may be seen in the top right panel of Fig.~\ref{FigRadProf}. Clearly the predicted profile is in excellent agreement with the simulation (dots correspond to individual gas particles in the simulation). 

The model temperature profile is shown in the bottom left panel of the same figure and is also in excellent agreement with the results from the simulation. We have verified that the model works equally well for other {\small RELHICs} of different masses. We have also verified that the overall properties of {\small RELHICs} are relatively insensitive to our choice of UV background. This may be seen in the inset of Figure~\ref{FigRadProf}, where the  dot-dashed curve shows the equilibrium temperature-density relation obtained if the intensity of the UV background is increased to match the $z=1$ HM01 spectrum. Using that relation leads to changes in the gas mass predicted for {\small RELHICs} smaller than $15\%$.

This analysis demonstrates that the gas in {\small RELHICs} is in hydrostatic equilibrium with the halo potential and that a simple model allows us to predict the mass, structural parameters, and radial profiles of the gas component of these ``dark'' minihalos accurately. In particular, we may use the same model to predict the neutral hydrogen content of {\small RELHICs}, an issue to which we turn next.

\begin{figure}
	\includegraphics[width=\columnwidth]{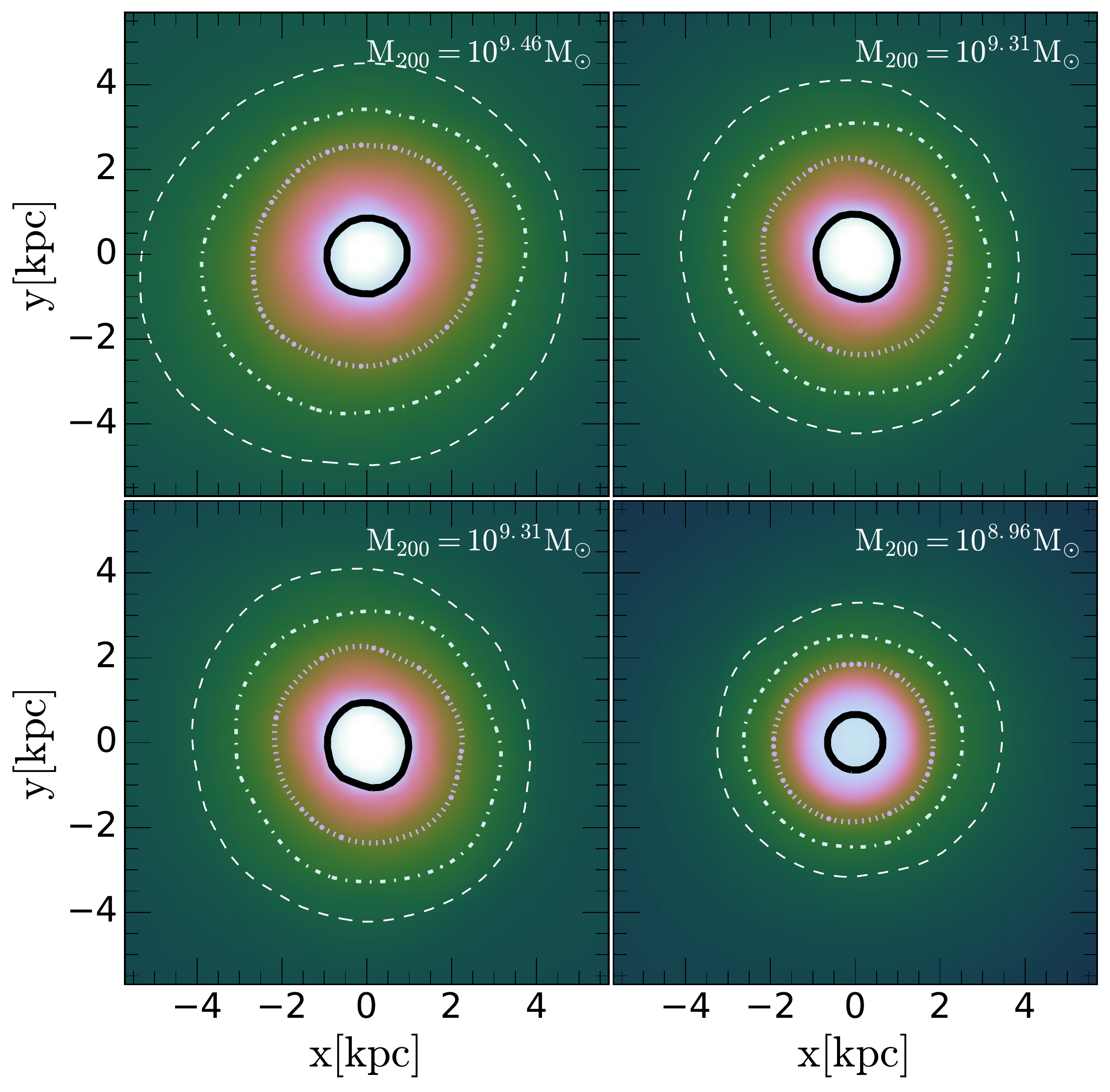}
    \caption{\ion{H}{I} column density maps of four {\small RELHICs}, chosen to have central column densities exceeding $10^{18}$ cm$^{-2}$. Contours correspond to column densities of $(5\times 10^{14}, 1\times 10^{15},4\times 10^{15}, 1\times 10^{18})$ atoms per $\rm cm^{2}$. Note that at relatively high column densities, {\small RELHICs} appear essentially round, with axis ratios $b/a>0.8$. This is consistent with the idea that {\small RELHICs} are in hydrostatic equilibrium in the potential of mildly triaxial dark matter halos.}
    \label{FigPhoto}
\end{figure}

\subsection{HI masses and radial profiles}
\label{SecHI}

The blue solid lines in Fig.~\ref{FigRadProf} show the density profiles of neutral hydrogen, derived using the fitting formula given in appendix A1 of~\cite{Rahmati2013}. This model uses a simple but accurate fit to the photoionization rates, obtained from radiative transfer simulations, where the scaling of the characteristic self-shielding density is taken from the analytic model of~\cite{Schaye2001}, and  computes neutral fractions as a function of density and temperature assuming ionization equilibrium. In the inner regions of the example {\small RELHIC} shown in Fig.~\ref{FigRadProf} the gas is dense and cold enough to be $\sim 100\%$ neutral; the neutral fraction drops rapidly from the center outwards. The \ion{H}{I} column density profile is shown in the bottom right panel of Fig.~\ref{FigRadProf}: the profile is quite steep, in part due to the onset of self-shielding in the model, and it drops from a well-defined central value of $10^{20}$ cm$^{-2}$ to $10^{18}$ cm$^{-2}$ at $\sim 1.25$ kpc from the centre\footnote{We compute the column density profiles by integrating the density profile along the line-of-sight within a sphere of radius $2\times r_{200}$.}. 

We show the gas density and \ion{H}{I} column density profiles in Fig.~\ref{FigModProf}, as a function of halo virial mass or, equivalently, as a function of the total gas mass. {\small RELHICs} have density profiles that vary in shape as the halo mass decreases, and central densities that correlate strongly with mass. Although the most massive {\small RELHICs} may reach central \ion{H}{I} column densities $\rm N_{\ion{H}{I},0}$ of $10^{21}$ cm$^{-2}$ these drop steeply with decreasing mass, dipping below $10^{15}$ cm$^{-2}$ for halos below $2.5 \times  10^9 \rm \ M_\odot$. This suggests that only the most massive {\small RELHICs} might be detectable in 21 cm surveys such as ALFALFA~\citep[e.g.,][]{Haynes2011}, which only reaches column densities exceeding $\sim 10^{18}$ cm$^{-2}$.

We provide further structural properties of the \ion{H}{I} component of {\small RELHICs} in Fig.~\ref{FigHI}, where we show the total \ion{H}{I} mass within $r_{200}$ as a function of central \ion{H}{I} column density and as a function of the total gas mass (left panels). The dashed lines show the results of the model described in Appendix~\ref{SecApp} ( Fig.~\ref{FigModProf}), which agree very well with the simulation results.  Clearly, neutral hydrogen makes up a very small fraction of the gaseous content of minihalos, confirming the expectations of the analytic models of \citet{Sternberg2002}: minihalos are essentially spheres of ionized gas in hydrostatic equilibrium and they have a small core of neutral hydrogen.

The bottom right-hand panel of Fig.~\ref{FigHI} shows several characteristic radii, where the {\small RELHIC} \ion{H}{I} column density drops from its central value to $10^{20}$, $10^{19}$,...,$10^{12}$ cm$^{-2}$, respectively (top to bottom).  The dashed lines indicate the results of the model profiles shown in Fig.~\ref{FigModProf}, whereas the open circles correspond to simulated {\small RELHICs}, grouped so that each is plotted at the radius where the column density drops by about one decade (or less) from the center. For example, the radii shown for {\small RELHICs} with central column densities between $10^{18}$ and $10^{19}$ cm$^{-2}$ is that where the column density drops to $10^{18}$ cm$^{-2}$, and so on. Note that, defined this way, most {\small RELHICs} within reach of current \ion{H}{I} surveys (i.e., $M_{\rm HI}> 10^4 \rm \  M_\odot$; $N_{\rm HI}>10^{18}$ cm$^{-2}$) are expected to be compact, sub-kpc systems \citep{Sternberg2002}.

{\small RELHICs} are near hydrostatic equilibrium, so the random bulk motions of the gas are quite small compared with the characteristic velocity dispersion of the halos they inhabit. This may be seen in the top right panel of Fig.~\ref{FigHI}, where we show, as a function of the total gas mass within $r_{200}$, the velocity dispersion of {\small RELHIC} gas particles, compared with the characteristic rms velocities of dark matter particles, $\sigma_{200}\approx V_{200}/\sqrt{2}$.
Typical bulk motions in {\small RELHICs} are below $5$ km/s in essentially all cases, well below $\sigma_{200}$, implying that the broadening of the 21 cm line should be mostly thermal.

Finally, we examine the morphologies of {\small RELHICs} in Fig.~\ref{FigPhoto}, which shows \ion{H}{I} column density maps for $4$ relatively massive {\small RELHICs} (see masses in figure legends). Drawing attention to the $10^{18}$ cm$^{-2}$ contour (black thick inner line), which corresponds to the sensitivity limit of surveys such as ALFALFA, we see that {\small RELHICs} would appear essentially round in such surveys. This is a direct consequence of the fact that the gas is in hydrostatic equilibrium in the dark halo potential. Indeed, although $\Lambda$CDM halos are intrinsically triaxial, the axis ratios of the potential are much less aspherical than those of the mass distribution \citep[see, e.g.,][]{Hayashi2007}.

\subsection{UCHVCs as Local Group RELHICs}
\label{SecObs}

We explore now the possibility that {\small RELHICs} might have been detected already in existing \ion{H}{I} surveys. Given their sizes and low \ion{H}{I} masses, we compare {\small RELHICs} with the population of Ultra Compact High Velocity Clouds (UCHVCs) first discussed by \citet{Giovanelli2010} in the context of the ALFALFA survey. These are identified as high signal-to-noise \ion{H}{I} sources with sizes less than $30'$, and velocities well outside the range expected for Galactic rotation. Note that $30'$ corresponds to $\sim 2$ kpc at a distance of $250$ kpc, so these sources might include sub-kpc {\small RELHICs} in the Local Group.

We begin by noting that, as shown in Fig.~\ref{FigSpatialDistribution}, {\small RELHICs} shun the region close to the Local Group barycentre and mainly populate the underdense regions of its outskirts. Indeed, we find no {\small RELHIC} within $500$ kpc of any of the two main LG galaxies in any of the three "high-resolution" volumes we have analysed. This has two important consequences; one is that, coupled with the low \ion{H}{I} masses expected of {\small RELHICs}, their \ion{H}{I} fluxes will be quite low, and another is that few {\small RELHICs} will have negative Galactocentric radial velocities, as most will still be expanding away outside the LG turnaround radius. 

\begin{figure}
	\includegraphics[scale=0.45]{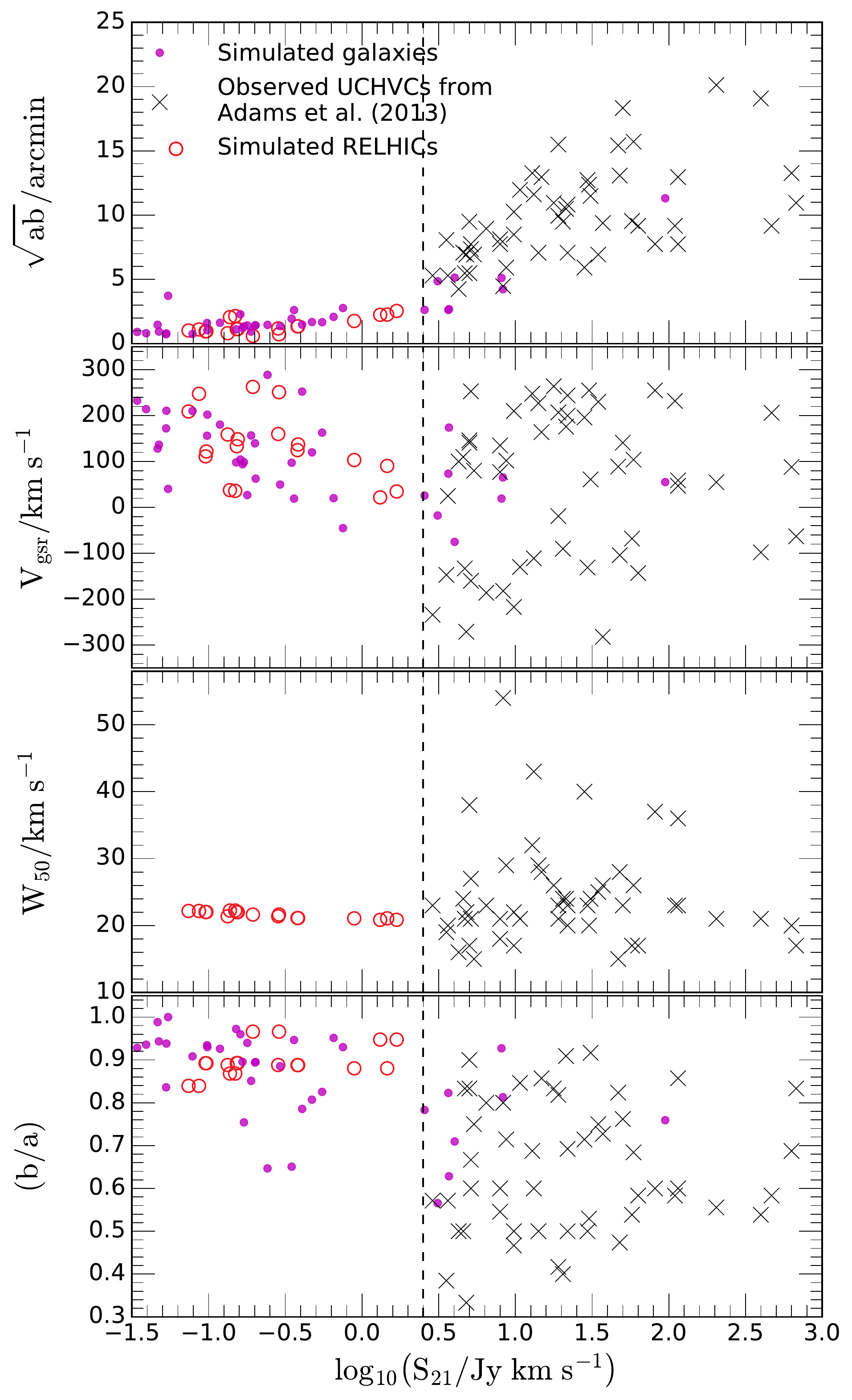}
    \caption{Simulated Local Group {\small RELHICs} (red open circles), and simulated LG dwarfs (magenta points) compared with the ALFALFA Ultra Compact High Velocity Clouds (UCHVCs, black crosses) from the compilation of \citet{Adams2013}. {\small RELHICs} are ``observed'' from the centre of the two primary galaxies, so that each {\small RELHIC} is shown twice. Only those with fluxes $S_{21}>0.1$ Jy km/s are shown. From top to bottom we show, as a function of $S_{21}$, the size of the $10^{18}$ cm$^{-2}$ \ion{H}{I} column density contour in arcmin; the Galactocentric radial velocity $V_{\rm gsr}$; the linewidth $W_{50}$; and the axis ratio $b/a$. Note that, in general, Local Group {\small RELHICs} are smaller in size, fainter, rounder, and more homogeneous in their linewidths than currently observed UCHVCs. {\small RELHICs} also have predominantly positive Galactocentric radial velocities, consistent with their large distances to the primaries ($d>500$ kpc). Magenta points indicate simulated galaxies with a non-zero stellar component ($M_{\rm str}< 10^6 \rm \ M_{\odot}$). These systems, in contrast, have properties resembling those of UCHVCs. Some are bigger in size, have higher fluxes, and exhibit a wider range of morphologies. Vertical dashed lines show the flux limit of the UCHVC compilation.
}
    \label{FigHVCs}
\end{figure}

We show this in Fig.~\ref{FigHVCs}, where we show, as a function of the \ion{H}{I} flux\footnote{We compute \ion{H}{I} fluxes using the total \ion{H}{I} mass within the virial radius of a {\small RELHIC}, $M_{\ion{H}{I}}$, and its distance to the LG primary galaxies expressed in Mpc, $d_{\rm Mpc}$: $M_{\rm HI}/{\rm M_\odot}=2.36 \times 10^5 \, S_{21}\, \left (d/\rm Mpc \right )^2$, with $S_{21}$ given in units of Jy km/s. Note that as we consider the two main galaxies of each simulated LG, every {\small RELHIC} is shown twice  in Fig.~\ref{FigHVCs}.}, $S_{21}$, in units of $\rm Jy \ km \ s^{-1}$, the \ion{H}{I} size of the {\small RELHIC}, defined as the mean radius $(\sqrt{ab})$, where $a$ and $b$ are the semiaxes of the best fitting ellipse to its $10^{18}$ cm$^{-2}$ isodensity contour (top panel),  Galactocentric radial velocity, $V_{\rm gsr}$ (second from top), the FWHM line broadening parameter\footnote{$W_{50}$ is computed by adding in quadrature the broadening due to the gas temperature and its bulk velocity dispersion. The former dominates, and is given by $T/$K$=21.8\, W_{50}^2$, with $W_{50}$ given in km/s.}, $W_{50}$ (third from top), and the axis ratio of the limiting \ion{H}{I} column density isocontour, $b/a$ (bottom panel).

Fig.~\ref{FigHVCs} compares simulated {\small RELHICs} (open red circles) and luminous simulated dwarfs ($M_{\rm str} < 10^6 \rm \  M_{\odot}$; magenta circles), with the $59$ UCHVCs catalogued by \citet{Adams2013} from ALFALFA data (see black crosses). The stellar mass limit roughly corresponds to that of Leo~P, which was discovered after follow-up imaging of an UCHVC ~\citep{Giovanelli2013}. Note that {\it all} {\small APOSTLE} {\small RELHICs} are just below the flux limit of the ALFALFA search, which is of $3$ Jy km/s (shown by the vertical dashed line). UCHVCs are also much more numerous and heterogeneous as a population than expected for {\small RELHICs}, which are rather small; fairly round ($b/a>0.8$); have a narrow dispersion of linewidths about $W_{50}\sim 20$ km/s; and are almost exclusively moving away from the Galaxy. UCHVCs are, however, comparable to simulated dwarfs, which reach higher fluxes, exhibit a wider range of morphologies, and are bigger than {\small RELHICs}. We do not analyse the distribution of $\rm W_{50}$ for our simulated dwarfs as their temperature is set by a an effective equation of state imposed to model the ISM, which is set to $10^4 \rm \ K$. Because of this, the HI fluxes estimated for dwarfs must be regarded as lower limits. 

We conclude that the \citet{Adams2013} UCHVC catalogue does not contain the star-free ``dark'' minihalos we associate with {\small RELHICs}, and that the properties of some UCHVCs might be consistent with very faint dwarf galaxies that have so far escaped detection in optical surveys. Indeed, as discussed in Sec.~\ref{SecIntro}, some UCHVCs have already been identified as low surface brightness galaxies, some in the Local Volume \citep{Sand2015} and some as far away as the Virgo Cluster \citep{Bellazzini2015a}.

\section{Summary and Conclusions}
\label{SecConc}

We have used the {\small APOSTLE} suite of cosmological hydrodynamical simulations of the Local Group to examine the gas content of $\Lambda$CDM minihalos. We focussed our analysis on systems that are free of stars in our highest-resolution runs, since in such systems the bound gas content at $z=0$ should only depend on the effects of the UV ionizing background and on the ram pressure stripping that affects minihalos as they travel through the cosmic web. 

``Dark'' minihalos (or, more precisely, systems with stellar mass $M_{\rm str}<10^4 \rm \ M_\odot$, the mass resolution limit of our simulations) split into two well-defined groupings: one where the mass of bound gas is set by the ionizing background and correlates tightly with the minihalo virial mass ({\small RELHICs}, for REionization Limited \ion{H}{I} Clouds), and another where there is little or no bound gas left within the halo after stripping by the cosmic web ({\small COSWEBs}, for COSmic WEb Stripped systems). The differentiation is thus mainly environmental; gas-free {\small COSWEBs} populate the  high-density regions near the luminous galaxies of the Local Group, where gas densities are high and cosmic web stripping is important, whereas the relatively gas-rich {\small RELHICs} inhabit the underdense outskirts. Few {\small RELHICs} are found within $500$ kpc of either the Milky Way or the M31 analogues in the simulations.

In terms of halo virial mass, the transition between luminous galaxies and dark systems like {\small RELHICs} and {\small COSWEBs} happens relatively quickly. Dark minihalos have masses that do not exceed $M_{200} \sim 10^{10} \rm \  M_\odot$; their fraction increase rapidly with decreasing mass, and they make up essentially all halos below $10^9 \rm \ M_\odot$. {\small RELHICs} make up most of the more massive dark minihalos; their abundance peaks at roughly $50\%$ for $M_{200}\sim 2\times 10^9 \rm \ M_\odot$. The {\small RELHIC} bound gas mass fraction decreases with decreasing mass; from $20\%$ of the universal baryon fraction  at $M_{200} \sim 5 \times 10^9 \rm \ M_\odot$ to  $0.3\%$ ($10^5 \rm \ M_\odot$, or ten particles in our highest-resolution runs) in $\sim 3\times 10^8 \rm \ M_\odot$ minihalos.

The gas component in {\small RELHICs} is in approximate hydrostatic equilibrium with the dark matter potential and in thermal equilibrium with the ionizing UV background. Their thermodynamic properties are therefore well understood, and their gas density and temperature profiles are in excellent agreement with a simple model where UV-heated gas is in thermal and hydrostatic equilibrium within NFW halos. Gas in {\small RELHICs} is nearly pristine in composition and nearly fully ionized, with small (sub-kpc) neutral hydrogen cores that span a large range of \ion{H}{I} masses and column densities. These cores have negligible Doppler broadening and nearly round morphologies. 

The most massive {\small RELHICs} have properties comparable to those of some Ultra Compact High Velocity Clouds (UCHVCs) but the bulk of the Local Group {\small RELHIC} population should have \ion{H}{I} fluxes just below $\sim 3$ Jy km/s, the limit of the ALFALFA UCHVC detection. 

Other differences between {\small RELHICs} and UCHVCs are the following: (i)  the sheer number of UCHVCs implies that most UCHVCs are not {\small RELHICs} (we expect fewer than 10 Local Group {\small RELHICs} with $S_{21}>0.1$ Jy km/s over the whole sky); (ii) {\small RELHICs} should mostly reside beyond $\sim 500$ kpc from the Milky Way, leading to low \ion{H}{I} fluxes ($<3$ Jy km/s), very small angular sizes ($<3'$), and predominantly positive Galactocentric radial velocities; (iii) {\small RELHICs} should be nearly round on the sky ($b/a>0.8$ at $10^{18}$ cm$^{-2}$) and (iv) have a very narrow distribution of thermally broadened line widths ($W_{50}\sim 20$ km/s). 

The small overlap in properties between UCHVCs and {\small RELHICs} suggest that the former are not part of the abundant dark minihalo population expected in the $\Lambda$CDM models. UCHVCs are either \ion{H}{I} ``debris'' in the Galactic halo, or else the \ion{H}{I} component of more massive halos, most of whom are expected to host a luminous stellar component as well. Further work is underway that aims to clarify the overall abundance of {\small RELHICs} in cosmological volumes; their contribution to the low-mass end of the \ion{H}{I} mass function; their relation to ultra faint galaxies; and the best strategies to detect them. Although {\small RELHICs} seem too faint to be a dominant source of \ion{H}{I} detections in extant or planned surveys, they may be easier to detect and study in absorption against the light of luminous background objects at moderate redshifts. {\small RELHICs} are a robust prediction of the $\Lambda$CDM paradigm so their detection and characterization would offer a unique opportunity to shed light onto the ``dark'' side of a cold dark matter-dominated universe.

\section{Acknowledgements}

We acknowledge useful discussions with John Cannon, Luke Leisman, Manolis Papastergis, Antonino Marasco and Tom Osterloo. We also thank the anonymous referee for valuable comments that helped to improve the paper. We have benefited from the following public Python packages: {\tt numpy} \citep{van2011numpy}, {\tt scipy} \citep{Jones2001}, {\tt matplotlib} \citep{Hunter2007}, {\tt Ipython} \citep{Perez2007} and {\tt py-sphviewer} \citep{Benitez-Llambay2015b}. RCA is a Royal Society University Research Fellow. This work was supported by the Science and Technology Facilities Council (gran number ST/L00075X/1) and the European Research Council (grant numbers GA 267291 "Cosmiway"). It was also partially supported by the Interuniversity Attraction Poles Programme initiated by the Belgian Science Policy Office ([AP P7/08 CHARM]), and by the European Research Council under the European Union's Seventh Framework Programme (FP7/2007-2013)/ERC grant agreement 278594-GasAround Galaxies and by the Netherlands Organisation for Scientific Research (NWO) through VICI grant 639.043.409.
This work used the DiRAC Data Centric system at Durham University,operated by the Institute for Computational Cosmology on behalf of the STFC DiRAC HPC Facility (www.dirac.ac.uk). This equipment was funded by
BIS National E-infrastructure capital grant ST/K00042X/1, STFC capital grants ST/H008519/1 and ST/K00087X/1, STFC DiRAC Operations grant ST/K003267/1 and Durham University. DiRAC is part of the National E-Infrastructure.

\bibliographystyle{mnras}
\bibliography{my_biblio}

\begin{thebibliography}{}
\makeatletter
\relax
\def\mn@urlcharsother{\let\do\@makeother \do\$\do\&\do\#\do\^\do\_\do\%\do\~}
\def\mn@doi{\begingroup\mn@urlcharsother \@ifnextchar [ {\mn@doi@}
  {\mn@doi@[]}}
\def\mn@doi@[#1]#2{\def\@tempa{#1}\ifx\@tempa\@empty \href
  {http://dx.doi.org/#2} {doi:#2}\else \href {http://dx.doi.org/#2} {#1}\fi
  \endgroup}
\def\mn@eprint#1#2{\mn@eprint@#1:#2::\@nil}
\def\mn@eprint@arXiv#1{\href {http://arxiv.org/abs/#1} {{\tt arXiv:#1}}}
\def\mn@eprint@dblp#1{\href {http://dblp.uni-trier.de/rec/bibtex/#1.xml}
  {dblp:#1}}
\def\mn@eprint@#1:#2:#3:#4\@nil{\def\@tempa {#1}\def\@tempb {#2}\def\@tempc
  {#3}\ifx \@tempc \@empty \let \@tempc \@tempb \let \@tempb \@tempa \fi \ifx
  \@tempb \@empty \def\@tempb {arXiv}\fi \@ifundefined
  {mn@eprint@\@tempb}{\@tempb:\@tempc}{\expandafter \expandafter \csname
  mn@eprint@\@tempb\endcsname \expandafter{\@tempc}}}

\bibitem[\protect\citeauthoryear{{Adams}, {Giovanelli}  \& {Haynes}}{{Adams}
  et~al.}{2013}]{Adams2013}
{Adams} E.~A.~K.,  {Giovanelli} R.,   {Haynes} M.~P.,  2013, \mn@doi [\apj]
  {10.1088/0004-637X/768/1/77}, \href
  {http://adsabs.harvard.edu/abs/2013ApJ...768...77A} {768, 77}

\bibitem[\protect\citeauthoryear{{Behroozi}, {Wechsler}  \&
  {Conroy}}{{Behroozi} et~al.}{2013}]{Behroozi2013}
{Behroozi} P.~S.,  {Wechsler} R.~H.,   {Conroy} C.,  2013, \mn@doi [\apj]
  {10.1088/0004-637X/770/1/57}, \href
  {http://adsabs.harvard.edu/abs/2013ApJ...770...57B} {770, 57}

\bibitem[\protect\citeauthoryear{{Bellazzini} et~al.,}{{Bellazzini}
  et~al.}{2015a}]{Bellazzini2015b}
{Bellazzini} M.,  et~al., 2015a, \mn@doi [\aap] {10.1051/0004-6361/201425262},
  \href {http://adsabs.harvard.edu/abs/2015A%26A...575A.126B} {575, A126}

\bibitem[\protect\citeauthoryear{{Bellazzini} et~al.,}{{Bellazzini}
  et~al.}{2015b}]{Bellazzini2015a}
{Bellazzini} M.,  et~al., 2015b, \mn@doi [\apjl] {10.1088/2041-8205/800/1/L15},
  \href {http://adsabs.harvard.edu/abs/2015ApJ...800L..15B} {800, L15}

\bibitem[\protect\citeauthoryear{Benitez-Llambay}{Benitez-Llambay}{2015}]{Benitez-Llambay2015b}
Benitez-Llambay A.,  2015, py-sphviewer: Py-SPHViewer v1.0.0,
  \mn@doi{10.5281/zenodo.21703}, \url {http://dx.doi.org/10.5281/zenodo.21703}

\bibitem[\protect\citeauthoryear{{Ben{\'{\i}}tez-Llambay}, {Navarro}, {Abadi},
  {Gottl{\"o}ber}, {Yepes}, {Hoffman}  \& {Steinmetz}}{{Ben{\'{\i}}tez-Llambay}
  et~al.}{2013}]{Benitez-Llambay2013}
{Ben{\'{\i}}tez-Llambay} A.,  {Navarro} J.~F.,  {Abadi} M.~G.,  {Gottl{\"o}ber}
  S.,  {Yepes} G.,  {Hoffman} Y.,   {Steinmetz} M.,  2013, \mn@doi [\apjl]
  {10.1088/2041-8205/763/2/L41}, \href
  {http://adsabs.harvard.edu/abs/2013ApJ...763L..41B} {763, L41}

\bibitem[\protect\citeauthoryear{{Blitz}, {Spergel}, {Teuben}, {Hartmann}  \&
  {Burton}}{{Blitz} et~al.}{1999}]{Blitz1999}
{Blitz} L.,  {Spergel} D.~N.,  {Teuben} P.~J.,  {Hartmann} D.,   {Burton}
  W.~B.,  1999, \mn@doi [\apj] {10.1086/306963}, \href
  {http://adsabs.harvard.edu/abs/1999ApJ...514..818B} {514, 818}

\bibitem[\protect\citeauthoryear{{Braun} \& {Burton}}{{Braun} \&
  {Burton}}{1999}]{Braun1999}
{Braun} R.,  {Burton} W.~B.,  1999, \aap, \href
  {http://adsabs.harvard.edu/abs/1999A%26A...341..437B} {341, 437}

\bibitem[\protect\citeauthoryear{{Bullock}, {Kravtsov}  \&
  {Weinberg}}{{Bullock} et~al.}{2000}]{Bullock2000}
{Bullock} J.~S.,  {Kravtsov} A.~V.,   {Weinberg} D.~H.,  2000, \mn@doi [\apj]
  {10.1086/309279}, \href {http://adsabs.harvard.edu/abs/2000ApJ...539..517B}
  {539, 517}

\bibitem[\protect\citeauthoryear{{Carlberg}}{{Carlberg}}{2009}]{Carlberg2009}
{Carlberg} R.~G.,  2009, \mn@doi [\apjl] {10.1088/0004-637X/705/2/L223}, \href
  {http://adsabs.harvard.edu/abs/2009ApJ...705L.223C} {705, L223}

\bibitem[\protect\citeauthoryear{{Charles} et~al.,}{{Charles}
  et~al.}{2016}]{Charles2016}
{Charles} E.,  et~al., 2016, \mn@doi [\physrep]
  {10.1016/j.physrep.2016.05.001}, \href
  {http://adsabs.harvard.edu/abs/2016PhR...636....1C} {636, 1}

\bibitem[\protect\citeauthoryear{{Crain} et~al.,}{{Crain}
  et~al.}{2015}]{Crain2015}
{Crain} R.~A.,  et~al., 2015, \mn@doi [\mnras] {10.1093/mnras/stv725}, \href
  {http://adsabs.harvard.edu/abs/2015MNRAS.450.1937C} {450, 1937}

\bibitem[\protect\citeauthoryear{{Dalal} \& {Kochanek}}{{Dalal} \&
  {Kochanek}}{2002}]{Dalal2002}
{Dalal} N.,  {Kochanek} C.~S.,  2002, \mn@doi [\apj] {10.1086/340303}, \href
  {http://adsabs.harvard.edu/abs/2002ApJ...572...25D} {572, 25}

\bibitem[\protect\citeauthoryear{{Diemand}, {Kuhlen}  \& {Madau}}{{Diemand}
  et~al.}{2007}]{Diemand2007}
{Diemand} J.,  {Kuhlen} M.,   {Madau} P.,  2007, \mn@doi [\apj]
  {10.1086/510736}, \href {http://adsabs.harvard.edu/abs/2007ApJ...657..262D}
  {657, 262}

\bibitem[\protect\citeauthoryear{{Dolag}, {Borgani}, {Murante}  \&
  {Springel}}{{Dolag} et~al.}{2009}]{Dolag2009}
{Dolag} K.,  {Borgani} S.,  {Murante} G.,   {Springel} V.,  2009, \mn@doi
  [\mnras] {10.1111/j.1365-2966.2009.15034.x}, \href
  {http://adsabs.harvard.edu/abs/2009MNRAS.399..497D} {399, 497}

\bibitem[\protect\citeauthoryear{{Fattahi} et~al.,}{{Fattahi}
  et~al.}{2016}]{Fattahi2016a}
{Fattahi} A.,  et~al., 2016, \mn@doi [\mnras] {10.1093/mnras/stv2970}, \href
  {http://adsabs.harvard.edu/abs/2016MNRAS.457..844F} {457, 844}

\bibitem[\protect\citeauthoryear{{Feldmann} \& {Spolyar}}{{Feldmann} \&
  {Spolyar}}{2015}]{Feldmann2015}
{Feldmann} R.,  {Spolyar} D.,  2015, \mn@doi [\mnras] {10.1093/mnras/stu2147},
  \href {http://adsabs.harvard.edu/abs/2015MNRAS.446.1000F} {446, 1000}

\bibitem[\protect\citeauthoryear{{Ferland}, {Korista}, {Verner}, {Ferguson},
  {Kingdon}  \& {Verner}}{{Ferland} et~al.}{1998}]{Ferland1998}
{Ferland} G.~J.,  {Korista} K.~T.,  {Verner} D.~A.,  {Ferguson} J.~W.,
  {Kingdon} J.~B.,   {Verner} E.~M.,  1998, \mn@doi [\pasp] {10.1086/316190},
  \href {http://adsabs.harvard.edu/abs/1998PASP..110..761F} {110, 761}

\bibitem[\protect\citeauthoryear{{Garrison-Kimmel}, {Boylan-Kolchin}, {Bullock}
   \& {Lee}}{{Garrison-Kimmel} et~al.}{2014}]{Garrison-Kimmel2014}
{Garrison-Kimmel} S.,  {Boylan-Kolchin} M.,  {Bullock} J.~S.,   {Lee} K.,
  2014, \mn@doi [\mnras] {10.1093/mnras/stt2377}, \href
  {http://adsabs.harvard.edu/abs/2014MNRAS.438.2578G} {438, 2578}

\bibitem[\protect\citeauthoryear{{Giovanelli} \& {Haynes}}{{Giovanelli} \&
  {Haynes}}{2016}]{Giovanelli2016}
{Giovanelli} R.,  {Haynes} M.~P.,  2016, \mn@doi [\aapr]
  {10.1007/s00159-015-0085-3}, \href
  {http://adsabs.harvard.edu/abs/2016A%26ARv..24....1G} {24, 1}

\bibitem[\protect\citeauthoryear{{Giovanelli}, {Haynes}, {Kent}  \&
  {Adams}}{{Giovanelli} et~al.}{2010}]{Giovanelli2010}
{Giovanelli} R.,  {Haynes} M.~P.,  {Kent} B.~R.,   {Adams} E.~A.~K.,  2010,
  \mn@doi [\apjl] {10.1088/2041-8205/708/1/L22}, \href
  {http://adsabs.harvard.edu/abs/2010ApJ...708L..22G} {708, L22}

\bibitem[\protect\citeauthoryear{{Giovanelli} et~al.,}{{Giovanelli}
  et~al.}{2013}]{Giovanelli2013}
{Giovanelli} R.,  et~al., 2013, \mn@doi [\aj] {10.1088/0004-6256/146/1/15},
  \href {http://adsabs.harvard.edu/abs/2013AJ....146...15G} {146, 15}

\bibitem[\protect\citeauthoryear{{Haardt} \& {Madau}}{{Haardt} \&
  {Madau}}{2001}]{Haardt2001}
{Haardt} F.,  {Madau} P.,  2001, in {Neumann} D.~M.,  {Tran} J.~T.~V.,  eds,
  Clusters of Galaxies and the High Redshift Universe Observed in X-rays.
  (\mn@eprint {} {astro-ph/0106018})

\bibitem[\protect\citeauthoryear{{Haehnelt}, {Rauch}  \&
  {Steinmetz}}{{Haehnelt} et~al.}{1996}]{Haehnelt1996}
{Haehnelt} M.~G.,  {Rauch} M.,   {Steinmetz} M.,  1996, \mn@doi [\mnras]
  {10.1093/mnras/283.3.1055}, \href
  {http://adsabs.harvard.edu/abs/1996MNRAS.283.1055H} {283, 1055}

\bibitem[\protect\citeauthoryear{{Hayashi}, {Navarro}  \& {Springel}}{{Hayashi}
  et~al.}{2007}]{Hayashi2007}
{Hayashi} E.,  {Navarro} J.~F.,   {Springel} V.,  2007, \mn@doi [\mnras]
  {10.1111/j.1365-2966.2007.11599.x}, \href
  {http://adsabs.harvard.edu/abs/2007MNRAS.377...50H} {377, 50}

\bibitem[\protect\citeauthoryear{{Haynes} et~al.,}{{Haynes}
  et~al.}{2011}]{Haynes2011}
{Haynes} M.~P.,  et~al., 2011, \mn@doi [\aj] {10.1088/0004-6256/142/5/170},
  \href {http://adsabs.harvard.edu/abs/2011AJ....142..170H} {142, 170}

\bibitem[\protect\citeauthoryear{{Hezaveh} et~al.,}{{Hezaveh}
  et~al.}{2016}]{Hezaveh2016}
{Hezaveh} Y.~D.,  et~al., 2016, \mn@doi [\apj] {10.3847/0004-637X/823/1/37},
  \href {http://adsabs.harvard.edu/abs/2016ApJ...823...37H} {823, 37}

\bibitem[\protect\citeauthoryear{Hunter}{Hunter}{2007}]{Hunter2007}
Hunter J.~D.,  2007, Computing In Science \& Engineering, 9, 90

\bibitem[\protect\citeauthoryear{{Ibata}, {Lewis}, {Irwin}  \& {Quinn}}{{Ibata}
  et~al.}{2002}]{Ibata2002}
{Ibata} R.~A.,  {Lewis} G.~F.,  {Irwin} M.~J.,   {Quinn} T.,  2002, \mn@doi
  [\mnras] {10.1046/j.1365-8711.2002.05358.x}, \href
  {http://adsabs.harvard.edu/abs/2002MNRAS.332..915I} {332, 915}

\bibitem[\protect\citeauthoryear{{Ikeuchi}}{{Ikeuchi}}{1986}]{Ikeuchi1986}
{Ikeuchi} S.,  1986, \mn@doi [\apss] {10.1007/BF00651178}, \href
  {http://adsabs.harvard.edu/abs/1986Ap%26SS.118..509I} {118, 509}

\bibitem[\protect\citeauthoryear{{Jenkins}}{{Jenkins}}{2013}]{Jenkins2013}
{Jenkins} A.,  2013, \mn@doi [\mnras] {10.1093/mnras/stt1154}, \href
  {http://adsabs.harvard.edu/abs/2013MNRAS.434.2094J} {434, 2094}

\bibitem[\protect\citeauthoryear{{Johnston}, {Spergel}  \& {Haydn}}{{Johnston}
  et~al.}{2002}]{Johnston2002}
{Johnston} K.~V.,  {Spergel} D.~N.,   {Haydn} C.,  2002, \mn@doi [\apj]
  {10.1086/339791}, \href {http://adsabs.harvard.edu/abs/2002ApJ...570..656J}
  {570, 656}

\bibitem[\protect\citeauthoryear{Jones, Oliphant, Peterson  et~al.}{Jones
  et~al.}{2001}]{Jones2001}
Jones E.,  Oliphant T.,  Peterson P.,   et~al., 2001, {SciPy}: Open source
  scientific tools for {Python}, \url {http://www.scipy.org/}

\bibitem[\protect\citeauthoryear{{Klypin}, {Kravtsov}, {Valenzuela}  \&
  {Prada}}{{Klypin} et~al.}{1999}]{Klypin1999}
{Klypin} A.,  {Kravtsov} A.~V.,  {Valenzuela} O.,   {Prada} F.,  1999, \mn@doi
  [\apj] {10.1086/307643}, \href
  {http://adsabs.harvard.edu/abs/1999ApJ...522...82K} {522, 82}

\bibitem[\protect\citeauthoryear{{Komatsu} et~al.,}{{Komatsu}
  et~al.}{2011}]{Komatsu2011}
{Komatsu} E.,  et~al., 2011, \mn@doi [\apjs] {10.1088/0067-0049/192/2/18},
  \href {http://adsabs.harvard.edu/abs/2011ApJS..192...18K} {192, 18}

\bibitem[\protect\citeauthoryear{{Mao} \& {Schneider}}{{Mao} \&
  {Schneider}}{1998}]{Mao1998}
{Mao} S.,  {Schneider} P.,  1998, \mn@doi [\mnras]
  {10.1046/j.1365-8711.1998.01319.x}, \href
  {http://adsabs.harvard.edu/abs/1998MNRAS.295..587M} {295, 587}

\bibitem[\protect\citeauthoryear{{Micha{\l}owski} et~al.,}{{Micha{\l}owski}
  et~al.}{2015}]{Michalowski2015}
{Micha{\l}owski} M.~J.,  et~al., 2015, \mn@doi [\aap]
  {10.1051/0004-6361/201526542}, \href
  {http://adsabs.harvard.edu/abs/2015A%26A...582A..78M} {582, A78}

\bibitem[\protect\citeauthoryear{{Moore}, {Ghigna}, {Governato}, {Lake},
  {Quinn}, {Stadel}  \& {Tozzi}}{{Moore} et~al.}{1999}]{Moore1999}
{Moore} B.,  {Ghigna} S.,  {Governato} F.,  {Lake} G.,  {Quinn} T.,  {Stadel}
  J.,   {Tozzi} P.,  1999, \mn@doi [\apjl] {10.1086/312287}, \href
  {http://adsabs.harvard.edu/abs/1999ApJ...524L..19M} {524, L19}

\bibitem[\protect\citeauthoryear{{Moster}, {Naab}  \& {White}}{{Moster}
  et~al.}{2013}]{Moster2013}
{Moster} B.~P.,  {Naab} T.,   {White} S.~D.~M.,  2013, \mn@doi [\mnras]
  {10.1093/mnras/sts261}, \href
  {http://adsabs.harvard.edu/abs/2013MNRAS.428.3121M} {428, 3121}

\bibitem[\protect\citeauthoryear{{Navarro}, {Frenk}  \& {White}}{{Navarro}
  et~al.}{1996}]{Navarro1996}
{Navarro} J.~F.,  {Frenk} C.~S.,   {White} S.~D.~M.,  1996, \mn@doi [\apj]
  {10.1086/177173}, \href {http://adsabs.harvard.edu/abs/1996ApJ...462..563N}
  {462, 563}

\bibitem[\protect\citeauthoryear{{Navarro}, {Frenk}  \& {White}}{{Navarro}
  et~al.}{1997}]{Navarro1997}
{Navarro} J.~F.,  {Frenk} C.~S.,   {White} S.~D.~M.,  1997, \apj, \href
  {http://adsabs.harvard.edu/abs/1997ApJ...490..493N} {490, 493}

\bibitem[\protect\citeauthoryear{P\'erez \& Granger}{P\'erez \&
  Granger}{2007}]{Perez2007}
P\'erez F.,  Granger B.~E.,  2007, \mn@doi [Computing in Science and
  Engineering] {10.1109/MCSE.2007.53}, 9, 21

\bibitem[\protect\citeauthoryear{{Rahmati}, {Pawlik}, {Raicevic}  \&
  {Schaye}}{{Rahmati} et~al.}{2013}]{Rahmati2013}
{Rahmati} A.,  {Pawlik} A.~H.,  {Raicevic} M.,   {Schaye} J.,  2013, \mn@doi
  [\mnras] {10.1093/mnras/stt066}, \href
  {http://adsabs.harvard.edu/abs/2013MNRAS.430.2427R} {430, 2427}

\bibitem[\protect\citeauthoryear{{Rauch}}{{Rauch}}{1998}]{Rauch1998}
{Rauch} M.,  1998, \mn@doi [\araa] {10.1146/annurev.astro.36.1.267}, \href
  {http://adsabs.harvard.edu/abs/1998ARA%26A..36..267R} {36, 267}

\bibitem[\protect\citeauthoryear{{Rees}}{{Rees}}{1986}]{Rees1986}
{Rees} M.~J.,  1986, \mn@doi [\mnras] {10.1093/mnras/218.1.25P}, \href
  {http://adsabs.harvard.edu/abs/1986MNRAS.218P..25R} {218, 25P}

\bibitem[\protect\citeauthoryear{{Sand} et~al.,}{{Sand}
  et~al.}{2015}]{Sand2015}
{Sand} D.~J.,  et~al., 2015, \mn@doi [\apj] {10.1088/0004-637X/806/1/95}, \href
  {http://adsabs.harvard.edu/abs/2015ApJ...806...95S} {806, 95}

\bibitem[\protect\citeauthoryear{{Saul} et~al.,}{{Saul}
  et~al.}{2012}]{Saul2012}
{Saul} D.~R.,  et~al., 2012, \mn@doi [\apj] {10.1088/0004-637X/758/1/44}, \href
  {http://adsabs.harvard.edu/abs/2012ApJ...758...44S} {758, 44}

\bibitem[\protect\citeauthoryear{{Sawala} et~al.,}{{Sawala}
  et~al.}{2016}]{Sawala2016}
{Sawala} T.,  et~al., 2016, \mn@doi [\mnras] {10.1093/mnras/stw145}, \href
  {http://adsabs.harvard.edu/abs/2016MNRAS.457.1931S} {457, 1931}

\bibitem[\protect\citeauthoryear{{Schaye}}{{Schaye}}{2001}]{Schaye2001}
{Schaye} J.,  2001, \mn@doi [\apjl] {10.1086/338106}, \href
  {http://adsabs.harvard.edu/abs/2001ApJ...562L..95S} {562, L95}

\bibitem[\protect\citeauthoryear{{Schaye}}{{Schaye}}{2004}]{Schaye2004}
{Schaye} J.,  2004, \mn@doi [\apj] {10.1086/421232}, \href
  {http://adsabs.harvard.edu/abs/2004ApJ...609..667S} {609, 667}

\bibitem[\protect\citeauthoryear{{Schaye} et~al.,}{{Schaye}
  et~al.}{2015}]{Schaye2015}
{Schaye} J.,  et~al., 2015, \mn@doi [\mnras] {10.1093/mnras/stu2058}, \href
  {http://adsabs.harvard.edu/abs/2015MNRAS.446..521S} {446, 521}

\bibitem[\protect\citeauthoryear{{Springel}}{{Springel}}{2005}]{Springel2005}
{Springel} V.,  2005, \mn@doi [\mnras] {10.1111/j.1365-2966.2005.09655.x},
  \href {http://adsabs.harvard.edu/abs/2005MNRAS.364.1105S} {364, 1105}

\bibitem[\protect\citeauthoryear{{Springel}, {White}, {Tormen}  \&
  {Kauffmann}}{{Springel} et~al.}{2001}]{Springel2001}
{Springel} V.,  {White} S.~D.~M.,  {Tormen} G.,   {Kauffmann} G.,  2001,
  \mn@doi [\mnras] {10.1046/j.1365-8711.2001.04912.x}, \href
  {http://adsabs.harvard.edu/abs/2001MNRAS.328..726S} {328, 726}

\bibitem[\protect\citeauthoryear{{Springel} et~al.,}{{Springel}
  et~al.}{2008}]{Springel2008}
{Springel} V.,  et~al., 2008, \mn@doi [\nat] {10.1038/nature07411}, \href
  {http://adsabs.harvard.edu/abs/2008Natur.456...73S} {456, 73}

\bibitem[\protect\citeauthoryear{{Sternberg}, {McKee}  \&
  {Wolfire}}{{Sternberg} et~al.}{2002}]{Sternberg2002}
{Sternberg} A.,  {McKee} C.~F.,   {Wolfire} M.~G.,  2002, \mn@doi [\apjs]
  {10.1086/343032}, \href {http://adsabs.harvard.edu/abs/2002ApJS..143..419S}
  {143, 419}

\bibitem[\protect\citeauthoryear{{Theuns}, {Leonard}, {Efstathiou}, {Pearce}
  \& {Thomas}}{{Theuns} et~al.}{1998}]{Theuns1998}
{Theuns} T.,  {Leonard} A.,  {Efstathiou} G.,  {Pearce} F.~R.,   {Thomas}
  P.~A.,  1998, \mn@doi [\mnras] {10.1046/j.1365-8711.1998.02040.x}, \href
  {http://adsabs.harvard.edu/abs/1998MNRAS.301..478T} {301, 478}

\bibitem[\protect\citeauthoryear{Van Der~Walt, Colbert  \& Varoquaux}{Van
  Der~Walt et~al.}{2011}]{van2011numpy}
Van Der~Walt S.,  Colbert S.~C.,   Varoquaux G.,  2011, Computing in Science \&
  Engineering, 13, 22

\bibitem[\protect\citeauthoryear{{Vegetti}, {Koopmans}, {Bolton}, {Treu}  \&
  {Gavazzi}}{{Vegetti} et~al.}{2010}]{Vegetti2010}
{Vegetti} S.,  {Koopmans} L.~V.~E.,  {Bolton} A.,  {Treu} T.,   {Gavazzi} R.,
  2010, \mn@doi [\mnras] {10.1111/j.1365-2966.2010.16865.x}, \href
  {http://adsabs.harvard.edu/abs/2010MNRAS.408.1969V} {408, 1969}

\bibitem[\protect\citeauthoryear{{White} \& {Rees}}{{White} \&
  {Rees}}{1978}]{White1978}
{White} S.~D.~M.,  {Rees} M.~J.,  1978, \mn@doi [\mnras]
  {10.1093/mnras/183.3.341}, \href
  {http://adsabs.harvard.edu/abs/1978MNRAS.183..341W} {183, 341}

\bibitem[\protect\citeauthoryear{{Wiersma}, {Schaye}  \& {Smith}}{{Wiersma}
  et~al.}{2009}]{Wiersma2009a}
{Wiersma} R.~P.~C.,  {Schaye} J.,   {Smith} B.~D.,  2009, \mn@doi [\mnras]
  {10.1111/j.1365-2966.2008.14191.x}, \href
  {http://cdsads.u-strasbg.fr/abs/2009MNRAS.393...99W} {393, 99}

\makeatother
\end{thebibliography}

\appendix
\section{Analytic model for RELHICs}
\label{SecApp}

We showed in Figure~\ref{FigRhoT} that the $n_{H}-T$ relation followed by gas particles in {\small RELHICs} effectively defines an equation of state, $P = P(\rho)$. Thus, heating and cooling processes couple the temperature of the gas to its density. We use this fact to derive a simple model that accounts for all the thermodynamic properties of RELIHCs. Our model relies on two main assumptions: 1) spherical symmetry and 2) hydrostatic equilibrium between the gas content of the halos and the gravitational potential, largely due to the dark matter distribution. 

We start by assuming a gaseous halo in hydrostatic equilibrium with its spherically symmetric potential, so that the pressure gradient is balanced by the halo gravitational acceleration:
\begin{equation}
\label{eq:HydroEqui1}
\displaystyle\frac{1}{\rho}\displaystyle\frac{dP}{d \tilde r} = - V_{200}^2 \displaystyle\frac{\tilde M(\tilde r)}{\tilde r^2} 
\end{equation}
\noindent where $V_{200}^2 = GM_{200}/r_{200} $ is the circular velocity of the halo at the virial radius $r_{200}$ and $\tilde M(\tilde r) = M(\tilde r)/M_{200}$ is the (normalized) enclosed mass within a sphere of radius $\tilde r = r/r_{200}$. For an ideal gas, the relation between pressure, density and temperature is given by:
\begin{equation}
P = \displaystyle\frac{\rho k_{\rm B} T}{\mu m_{\rm p}} ,
\end{equation}
\noindent where $k_{\rm B}$ is the Boltzmann constant, $\mu$ is the gas mean molecular weight, $m_{\rm p}$ is the proton mass and $\gamma$ is the adiabatic index or ratio of specific heats. Throughout this paper we use $\gamma = 5/3$ and $\mu = 0.6$, although allowing $\mu$ to vary might lead to an improvement of the model. 

Equation~\ref{eq:HydroEqui1} can be solved if we know the pressure at a particular radius, or equivalently, the density and the temperature at that radius. However, for our particular purposes the gas temperature is defined by its density, and thus the pressure is set by the gas density only, so that Eq.~\ref{eq:HydroEqui1} can be rewritten as:
\begin{equation}
\label{eq:HydroEqui2}
\left (  \frac{T}{\rho} +  \frac{dT}{d\rho}\right ) d\rho = - 2 T_{200} \displaystyle\frac{\tilde M(\tilde r)}{\tilde r^2} d\tilde r,
\end{equation}
\noindent where we have introduced the virial temperature
\begin{equation}
T_{200} = \displaystyle\frac{\mu m_{\rm p}}{2 k_{\rm B}}V_{200}^2 \sim 10^4 {\rm K} \left ( \displaystyle\frac{V_{200}}{17 \rm km \ s^{-1}}\right )^2.
\end{equation}
Assuming that the halo potential is largely due to the underlying dark matter distribution, we can model it with a NFW mass profile, so that the acceleration profile is:
\begin{equation}
\label{Eq:AccProfile}
\displaystyle\frac{\tilde M(\tilde r)}{\tilde r^2} d\tilde r = \displaystyle\frac{1}{\tilde r^2}  \displaystyle\frac{\ln(1+c\tilde r) - c\tilde r/(1+c\tilde r)}{\ln(1+c)-c/(1+c)} d\tilde r,
\end{equation}
\noindent where $c$ is the concentration parameter. 

We now study the asymptotic behaviour of the model. At small radii, ($ \tilde{r} << 1$), the gas density converges as $(1+c\tilde r)^{-1}$. In fact, the integral of the acceleration profile is: 
\begin{equation}
\displaystyle\int \displaystyle\frac{\tilde M (\tilde r')}{\tilde r'^2} d\tilde r' \propto \displaystyle\frac{\ln(1+c\tilde r)}{\tilde r}
\end{equation} 
For larger distances, $(\tilde r>>1)$, the acceleration profile vanishes, thus implying that the density eventually reaches a constant value. Moreover, for a sufficiently isolated halo, the density will reach the mean gas density of the Universe $\bar \rho$ at larger radii. This simple and theoretically motivated boundary condition is in fact the only free parameter of the model, and enables us to predict the gas mass within $r_{200}$, as a function of $M_{200}$, as shown by the purple solid line in Figure~\ref{FigMasses}, and it is enough to derive all other properties shown in Figures~\ref{FigModProf}, \ref{FigHI}, \ref{FigPhoto} and \ref{FigHVCs}.  

\begin{table}
	\centering
	\caption{Values of the temperature-density relation followed by RELHICs.}
	\begin{tabular}{llll} 
		\hline
		$\log_{10} \left (n_{\rm H} / \rm cm^{-3} \right )$ & $\log_{10} \left ( T / \rm \ K \right ) $ & $\log_{10} \left (n_{\rm H} / \rm cm^{-3} \right )$ & $\log_{10} \left ( T / \rm \ K \right ) $ \\
		\hline
	-8.0	& 2.91 & -3.8	& 4.38 \\
	-7.8	& 3.02 & -3.6	& 4.32 \\
	-7.6	& 3.13 & -3.4	& 4.28 \\
	-7.4	& 3.24 & -3.2	& 4.24 \\
	-7.2	& 3.36 & -3.0	& 4.20 \\
	-7.0	& 3.47 & -2.8	& 4.17 \\
	-6.8	& 3.59 & -2.6	& 4.14 \\
	-6.6	& 3.70 & -2.4	& 4.12 \\
	-6.4	& 3.81 & -2.2	& 4.10 \\
	-6.2	& 3.93 & -2.0	& 4.08 \\
	-6.0	& 4.04 & -1.8	& 4.06 \\
	-5.8	& 4.15 & -1.6	& 4.04 \\
	-5.6	& 4.25 & -1.4	& 4.03 \\
	-5.4	& 4.35 & -1.2	& 4.01 \\
	-5.2	& 4.44 & -1.0	& 4.00 \\
	-5.0	& 4.51 & -0.8	& 3.99 \\
	-4.8	& 4.54 & -0.6	& 3.97 \\
	-4.6	& 4.54 & -0.4	& 3.96 \\
	-4.4	& 4.51 & -0.2	& 3.95 \\
	-4.2	& 4.49 &  0.0   & 3.94 \\
	-4.0	& 4.43 &           &       \\
		\hline		
	\end{tabular}
	\label{Tab:table_fit}
\end{table}

In practice, we solve Eq.~\ref{eq:HydroEqui2} numerically, imposing the relation between temperature and density shown by the red-dashed line in Figure~\ref{FigRhoT} (given in Table~\ref{Tab:table_fit} for completeness), and fitting a NFW acceleration profile to the mass distribution of the simulated halos. In brief, we solve:
\begin{eqnarray}
F(\rho) &=& \displaystyle\int_{\bar \rho}^{\rho} \left. \left ( \displaystyle\frac{T}{\rho} + \displaystyle\frac{dT}{d\rho} \right ) \right|_{\rho'} d\rho' \\
G(r) &=& -2 T_{200} \displaystyle\int_{\infty}^{r} \displaystyle\frac{\tilde M(\tilde r')}{\tilde r'^2} d\tilde r',
\end{eqnarray}
\noindent and the spatial dependence of density is obtained by inverting F:
\begin{equation}
\rho(r) = F^{-1} \left [ G(r) \right ].
\end{equation}

For low-density gas ($n_{\rm H} < 10^{-4.8} \ \rm cm^{-3}$), $T(\rho)$ is well approximated by a power law, and an analytical solution can be given. In fact, assuming 
\begin{equation}
T(\rho) = T_{0} \left (\displaystyle\frac{\rho}{\rho_0} \right )^{\gamma_0},
\end{equation}

\noindent where $T_{0} \sim 10^4 \rm \ K$, $\left ( \rho_0/m_{\rm p} \right ) \sim 10^{-6} \rm \ cm^{-3}$ and $\gamma_{0} \sim 0.54$, it is straightforward to integrate Eq.~\ref{eq:HydroEqui2} to obtain the gas density profile:
\begin{equation}
\rho (\tilde r) = \bar{\rho} \left \{ \displaystyle\frac{2\gamma_0}{(1+\gamma_0)} \left ( \displaystyle\frac{T_{200}}{T_{0}}\right ) \left ( \displaystyle\frac{\rho_0}{\bar{\rho}}\right )^{\gamma_0} \displaystyle\frac{\ln(1+c\tilde r)}{\left [ \ln ( 1 + c ) - \frac{c}{1+c} \right ] \tilde r} + 1 \right \}^{1/\gamma_0}.
\end{equation}

\section{Star formation density threshold}
\label{SecApp2}

\begin{figure}
	\includegraphics[width=\columnwidth]{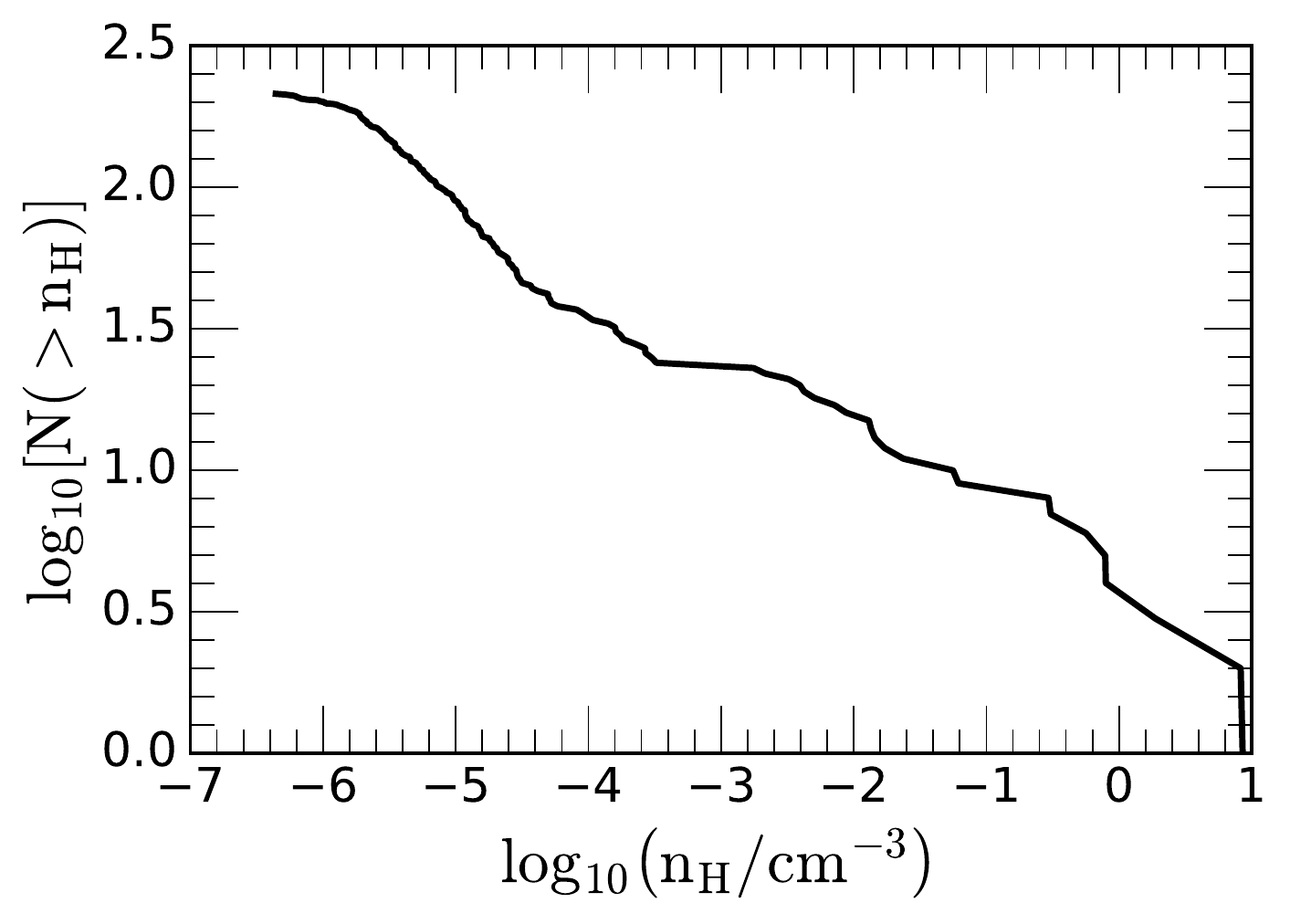}
    \caption{Cumulative number of {\small RELHICs} as a function of their central density, $n_{\rm H}$. Note that by construction, {\it none} of the them reach densities greater than $10 \rm \ cm^{-3}$, as otherwise they would form stars. If we lower the density threshold for forming stars by a factor of $1000$, to $n_{\rm H, th} = 0.01 \rm \ cm^{-3}$, we would remove 14 {\small RELHICs}. We note, however, that because ISM (gas with $n_{\rm H} \ge 0.1 \rm \ cm^{-3}$) is modelled by an effective equation of state, we have already removed $9$ {\small RELHICs} from analysis.}
    \label{FigApp}
\end{figure}

Star formation typically occurs when gas is able to develop a cold phase. We note, however, that our simulations lack the physics needed to simulate the transition between the warm and cold phase self-consistently, so that gas particles are eligible to form stars once they reach densities above a given threshold, $n_{\rm H, th}$, at a temperature of $\sim 10^4 \rm \ K$. In {\small APOSTLE} we use a density threshold that depends on metallicity, as proposed by~\citet{Schaye2004}:
\begin{equation}
\label{Eq:StarFormation}
n_{\rm H,th} (Z) = {\rm min} \left [ 0.1 \left ( \displaystyle\frac{Z}{0.002} \right )^{-0.64}, 10 \right ] \rm cm^{-3},
\end{equation}
which takes into account that the transition between warm, neutral phase to a cold, molecular one occurs at lower densities in more metal-rich gas. Eq.~\ref{Eq:StarFormation} is strictly valid for $Z>10^{-4} \rm \ Z_{\odot}$, and thus the maximum threshold, $n_{\rm H,th} = 10 \rm \ cm^{-3}$, valid for the extremely low metallicity {\small RELHICs}, is somehow arbitrary. We note, however, that in low-metallicity systems, gas may become cold and form stars without developing a molecular phase~\citep[see, e.g.,][]{Michalowski2015}, a mechanism that cannot be captured in APOSTLE. We can study, however, the impact of choosing a different density threshold for star formation in low-metallicity systems. We quantify this in Figure~\ref{FigApp}, where the cumulative number of {\small RELHICs} as a function of their central density is shown. By construction, {\it none} of the {\small RELHICs} reach central densities above $10 \rm \ cm^{-3}$. There are $\sim 9$ with densities greater than $0.1 \rm cm^{-3}$, which were already excluded from analysis (see Sec.~\ref{SecRhoT}), and $\sim 14$ with densities above $0.01 \rm cm^{-3}$. We conclude that reducing the threshold density for star formation by a factor of $1000$ respect to the current value would lead to a removal of $\sim 5$ {\small RELHICs} included in our current analysis.

\end{document}